\documentclass[useAMS,usenatbib]{mn2e}
\usepackage[cp1251]{inputenc}
\usepackage{amsmath}
\usepackage{amssymb}
\usepackage{euscript}
\usepackage[dvips]{graphicx}
\usepackage[T1]{fontenc}
\usepackage{aecompl}
\usepackage{epsfig}
\usepackage{ulem}
\usepackage[usenames,dvipsnames]{color}

\def\la{\;\raise0.3ex\hbox{$<$\kern-0.75em\raise-1.1ex\hbox{$\sim$}}\;}
\def\ga{\;\raise0.3ex\hbox{$>$\kern-0.75em\raise-1.1ex\hbox{$\sim$}}\;}

\title[Partial covering of Q\,0528$-$250 by intervening H$_2$ clouds.]
{Partial covering of emission regions of Q\,0528$-$250 by intervening H$_2$ clouds.}

\author[Klimenko et al.]{V.V.~Klimenko$^{1,2}$, S.A.~Balashev$^{1,2}$,
A.V.~Ivanchik$^{1,2}$, C.~Ledoux$^4$,
\newauthor P.~Noterdaeme$^3$, P.~Petitjean$^3$, R.~Srianand$^5$, D.A.~Varshalovich$^{1,2}$
\\
\vspace{4pt}\\
$^1$Ioffe Physical-Technical Institute of RAS, {Polyteknicheskaya 26}, 194021 Saint-Petersburg, Russia\\
$^2$St.-Petersburg State Polytechnical University, {Polyteknicheskaya 29}, 195251 Saint-Petersburg, Russia \\
$^3$Universit\'e Pierre et Marie-Curie, Institut d'Astrophysique de Paris, CNRS-UMR7095, 98bis boulevard Arago, 75014
Paris, France \\
$^4$European Southern Observatory, Alonso de C\'ordova 3107, Casilla
19001, Vitacura, Santiago 19, Chile\\
$^5$ Inter-University Centre for Astronomy and Astrophysics, Post~Bag~4, Ganesh~Khind, Pune 411~007, India \\
}
\begin{document}

\date{Accepted 15.12.2014 Received 08.10.2014}

\pagerange{\pageref{firstpage}--\pageref{lastpage}} \pubyear{2015}

\maketitle

\label{firstpage}

\begin{abstract}
We present an analysis of the molecular hydrogen absorption system at $z_{abs}=2.811$ in the spectrum of
the blazar Q\,0528$-$250. We demonstrate that the molecular cloud does not cover the background source
completely. The partial coverage reveals itself as a residual flux in the bottom of saturated H$_2$
absorption lines. This amounts to about $(2.22\pm0.54)\%$ of the continuum and does not depend on the
wavelength. This value is small and it explains why this effect has not been detected in previous studies
of this quasar spectrum. However, it is robustly detected and significantly higher than the zero flux
level in the bottom of saturated lines of the Ly$\alpha$ forest, $(-0.21\pm0.22)$ per cent. The presence
of the residual flux could be caused by unresolved quasar multicomponents, by light scattered by dust,
and/or by jet-cloud interaction. The H$_2$ absorption system is very well described by a two-component
model without inclusion of additional components when we take partial coverage into account. The derived
total column densities  in the H$_2$ absorption components A and B are $\log N({\rm H}_2)[{\rm
cm^{-2}}]=18.10\pm0.02$ and 17.82\,$\pm$\,0.02, respectively. HD molecules are present only in component
B. Given the column density, $\log N{\rm(HD)}=13.33\,\pm\,0.02$, we find $N({\rm{HD}})/\,2
N(\rm{H_2})\,=\,(1.48\,\pm\,0.10)\,\times\,10^{-5}$, significantly lower than previous estimations. We
argue that it is crucial to take into account the partial coverage effects in any analysis of H$_2$
bearing absorption systems, in particular when studying the physical state of high-redshift interstellar
medium.
\end{abstract}


\section{Introduction}
\label{introduction}
\noindent

As a result of their cosmological distances quasars (QSOs) appear as point-like objects. Various studies
have aimed to explore the detailed inner structure of quasars, which is unresolved even for low-redshift
active galactic nuclei (AGNs), because of their remote distances and sub-parsec scales of their emission
regions. In the standard AGN paradigm the central region is divided into an accretion disk, a dusty-torus,
a jet, a broad line region (BLR), and a narrow line region (NLR). Each of these regions contribute
differently to the AGN emission spectrum.

Direct imaging of the spatial structure of AGNs is possible with current instruments mainly probing longer
scales. To date, several interferometric studies of the central engine of the brightest AGNs (e.g.
\citealt{Jaffe2004, Tristam2013, Lopez2014}) have revealed the existence of a hot, parsec-scale disk that
is surrounded by warm dust extended in the polar direction. In the optical band the geometry of the
emission line region is investigated by indirect methods. Reverberation mapping establishes the
relationship between the size and the luminosity of the BLR and yields a typical BLR size of ${\rm R}_{\rm
BLR}\sim$\,0.2\,pc \citep{Kaspi2007, Chelouche2012} for high redshift luminous quasars. Differential
microlensing allows for a constraint on the accretion disk size $\la 3\times10^{-3}$\,pc
\citep{Blackburne2011, Vicente2012} and for an estimation of  the size of the BLR $\sim0.1$ pc
\citep{Sluse2011}. The observations of gamma-ray emission constrain the size of a jet constituent to a few
parsecs \citep{Abdo2010}.

Another estimate of the size of the AGN emitting regions comes from constraints derived from covering
factor analysis of intervening H$_2$ bearing clouds which happen to cover the background source only
partially. Analysis of the partial coverage of Q\,1232$+$082 by a molecular hydrogen absorption cloud
allowed \citet{Balashev2011} to estimate the size of the C\,{\sc IV} BLR, R$_{\mbox{\small C\,{\sc
IV}}}\sim\,0.16^{+0.08}_{-0.11}$\,pc.

Molecular hydrogen absorption systems, a subset of damped Ly$\alpha$ systems (DLAs) and sub-damped
Ly$\alpha$ systems (sub-DLAs), reveal diffuse and translucent interstellar clouds in high redshift
intervening galaxies \citep{Noterdaeme2008}. An analysis of H$_2$ absorption systems allows for examining
the physical conditions of diffuse clouds in distant galaxies \citep{Srianand2005, Noterdaeme2007}. It has
been shown that the gas is a part of the cold neutral medium with comparatively low kinetic temperature
($T\sim 50-500\,$K) and high densities ($n_{\rm H}$~$>$~10~cm$^{-3}$), thus compact sizes ($l \la 1\,$pc).
Comparison of 21 cm and H$_2$ absorptions suggests that the H$_2$ absorption originates from a compact gas
that probably contains only a small fraction of H\,{\sc i} measured along the line of sight
\citep{Srianand2012}. These systems are important instruments for the analysis of several cosmological
problems, as follows. (i) The discovery of HD/H$_2$ clouds at high redshift \citep{Varshalovich2001}
provides an independent way to estimate the primordial deuterium abundance (D/H) and therefore the
relative baryon density of the Universe $\Omega_{\rm b}$ which is one of the key cosmological parameters
\citep{Balashev2010, Ivanchik2010}. (ii) The comparison of H$_2$ wavelengths observed in QSO spectra with
laboratory ones (just for this quasar Q\,0528$-$250, \citet{Varshalovich1993}, \citet{Cowie1995},
\citet{Potekhin1998}, \citet{Ubachs2004}, \citet{King2011}) allows us to test the possible cosmological
variation of the proton-to-electron mass ratio $\mu=m_p/m_e$. Because the de-composition of H$_2$
absorptions into several components is crucial for studies of the fundamental constant variability
problem, we should pay attention to the partial coverage effect. It is known that taking into account the
partial coverage effects, the physical model of the absorption system differs (see
\citealt{Balashev2011}). (iii) The interpretation of the relative populations of C\,{\sc i} fine-structure
excitation levels and CO rotational levels (\citealt{Srianand2000,Noterdaeme2011}) allows us to measure
the temperature $T_{\rm CMB}(z)$ of the cosmic microwave background radiation at high redshift.

Here, we argue that it is necessary to take into account the partial coverage of quasar emission regions
by a compact intervening H$_2$ cloud in order to derive a robust fit of the absorption lines. If this
effect is not taken into account, column densities can be underestimated by a factor of up to two orders
of magnitude. The first case of such an analysis has been presented by \citealt{Balashev2011} for
Q\,1232$+$082. The second case of partial coverage has been detected by \cite{Albornoz2014} for H$_2$
bearing cloud towards the quasar Q\,0643$-$504. The third case of partial coverage for the H$_2$ cloud at
$z_{\rm abs}=2.811$ in the spectrum of Q\,0528$-$2508 is presented in this study. We analyse a new
spectrum and detect residual flux in the bottom of saturated H$_2$ lines (${\rm J=0}$ and ${\rm J=1}$
levels). In case this flux is not taken into account, saturated lines yield large $\chi^2$ values and a
multicomponent model is used instead (e.g. \citealt{King2011}).

The remainder of this paper is organized as follows. A brief description of the data is given in
Section\,\ref{data}. The principles of partial coverage are described in Section\,\ref{pc}. In
Section\,\ref{H2analysis}, we present the analysis of the H$_2$ absorption system, accounting for partial
coverage . The HD molecular lines are explored in Section\,\ref{HD}. The results are discussed in
Section\,\ref{discussion}, and we give a brief conclusion in Section\,\ref{conclusion}.

\section{Data}
\label{data}
\noindent

The molecular hydrogen was identified for the first time at high redshift in the very spectrum of
Q\,0528$-$2508 \citep{Levshakov1985}. This quasar was observed many times, in particular during the period
between 2001 and 2009 using both spectroscopic arms of Ultraviolet and Visual Echelle Spectrograph (UVES)
of the Very Large Telescope (VLT); for a description of the instrument, see \citet{Dekker2000}. The log of
the observations used in our work is shown in Table~\ref{table_obs}. These observations relate to four
programmes, three of which were carried out in 2001--2002: 66.A-0594(A) (PI: Molaro), 68.A-0600(A) (PI:
Ledoux), and 68.A-0106(A) (PI: Petitjean). The instrument settings used during these observations were a
1-arcsec slit and 2x2 CCD pixel binning in both arms, resulting in a resolving power of R$\sim$45000 in
the blue and R$\sim$43000 in the red. There was no ThAr lamp calibration attached to each of the
exposures. An additional series of observations was performed in 2008-2009 under programme 082.A-0087(A)
(PI: Ubachs). The settings for that programme were a 0.8-arcsec slit in the blue arm and a 0.7-arcsec slit
in the red. The 2x2 CCD pixel binning was also used at that time. This resulted in a resolving power of
R$\sim $60000 in the blue and R$\sim $56000 in the red. Because the goal of that programme was to set a
limit on the variation of $\mu$, ThAr lamp calibrations were also taken immediately after each
observation.

The data presented in Table 1 were reduced using the UVES Common Pipeline Library (CPL) data reduction
pipeline release 4.9.5 using the optimal extraction method\footnote{see the UVES pipeline user manual
available for download at ftp://ftp.eso.org/pub/dfs/pipelines/uves/uves-pipeline-manual-22.8.pdf}.  The
inter-order background (scattered light inside the instrument) was carefully subtracted in both the
flat-field frames and the science exposures. Linear spline interpolation was used to produce a
two-dimensional background image, which was subsequently smoothed using an average boxcar. Fourth-order
polynomials were used to find the dispersion solutions. However, the errors only reflect the calibration
error at the observed wavelengths of the ThAr lines used for wavelength calibration. All the spectra were
corrected for the motion of the observatory around the barycentre of the Solar system and then converted
to vacuum wavelengths. These spectra were interpolated into a common wavelength array and generated the
weighted-mean combined spectrum using the inverse squares of errors as weights. All the available
exposures were utilized to increase the signal-to-noise ratio up to $\sim$60 per pixel in the wavelength
range of the H$_2$ absorption lines (at $z=2.811$). As shown below, this allows us to detect and study the
effect of partial coverage in details.

\begin{table}
    \caption{Log of the observations}
    \begin{center}
     \begin{tabular}{|c|c|c|c|c|}
            \hline
            No. & UT Date & Program ID  & Exposure & Slit \\
                &         &             &    (sec) & (arcsec) \\
            \hline
            1  & 03.02.2001 & 66.A-0594(A) & 1$\times$5655 & 1.0  \\
            2  & 04.02.2001 & 66.A-0594(A) & 2$\times$5655 & 1.0  \\
            3  & 05.02.2001 & 66.A-0594(A) & 1$\times$5655 & 1.0  \\
            4  & 07.02.2001 & 66.A-0594(A) & 1$\times$5655 & 1.0  \\
            5  & 13.02.2001 & 66.A-0594(A) & 1$\times$5655 & 1.0  \\
            6  & 13.03.2001 & 66.A-0594(A) & 1$\times$5655 & 1.0  \\
                  7  & 15.03.2001 & 66.A-0594(A) & 1$\times$5655 & 1.0  \\
                  8  & 17.10.2001 & 68.A-0600(A) & 1$\times$3600 & 1.0  \\
                  9 & 18.10.2001 & 68.A-0600(A) & 2$\times$3600 & 1.0  \\
            10 & 08.01.2002 & 68.A-0106(A) & 2$\times$3600 & 1.0  \\
            11 & 09.01.2002 & 68.A-0106(A) & 2$\times$3600 & 1.0  \\
            12 & 10.01.2002 & 68.A-0106(A) & 2$\times$3600 & 1.0  \\
            13 & 23.11.2008 & 082.A-0087(A) & 2$\times$2900 & 0.8-0.7 \\
            14 & 25.11.2008 & 082.A-0087(A) & 1$\times$2900 & 0.8-0.7 \\
            15 & 23.12.2008 & 082.A-0087(A) & 4$\times$2900 & 0.8-0.7 \\
            16 & 25.01.2009 & 082.A-0087(A) & 1$\times$2900 & 0.8-0.7 \\
            17 & 26.01.2009 & 082.A-0087(A) & 1$\times$2900 & 0.8-0.7 \\
            18 & 26.02.2009 & 082.A-0087(A) & 1$\times$2900 & 0.8-0.7 \\
            \hline
            Total & \multicolumn{2}{c}{} & 108450 & \\
            \hline
     \end{tabular}
     \label{table_obs}
     \end{center}
\end{table}

\section{Effect of partial covering}
\label{pc}
\noindent

Partial covering implies that only a part of the background source is covered by the absorbing cloud.
Mainly this can be a consequence of the absorbing cloud size being comparable to, or even smaller than,
the projected extent of the background source.
\begin{figure}
\begin{center}
        \includegraphics[width=0.47\textwidth]{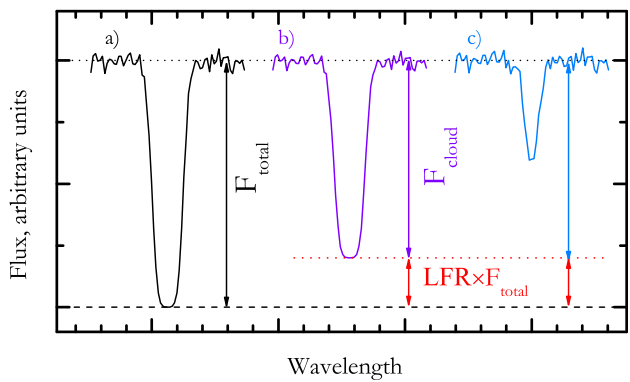}
        \caption{\rm An illustration of the effect of partial coverage on the absorption-line profiles.
        Different panels show (a) a highly saturated line with total coverage, $f_{\rm c}=1\,$,\,
        (b) a highly saturated line with partial coverage, $f_{\rm c}=0.8\,$\, and
        (c) a partially saturated line with the same partial coverage, $f_{\rm c}=0.8\,$. The line flux residual
        (LFR) is the fraction of the QSO flux that is not intercepted by the cloud. It can be easily derived from
        the spectral analysis in case (b), but it is not straightforward to detect in case (c).}
        \label{Def_CF}
\end{center}
\end{figure}

\begin{figure*}
\begin{center}
        \includegraphics[width=0.95\textwidth]{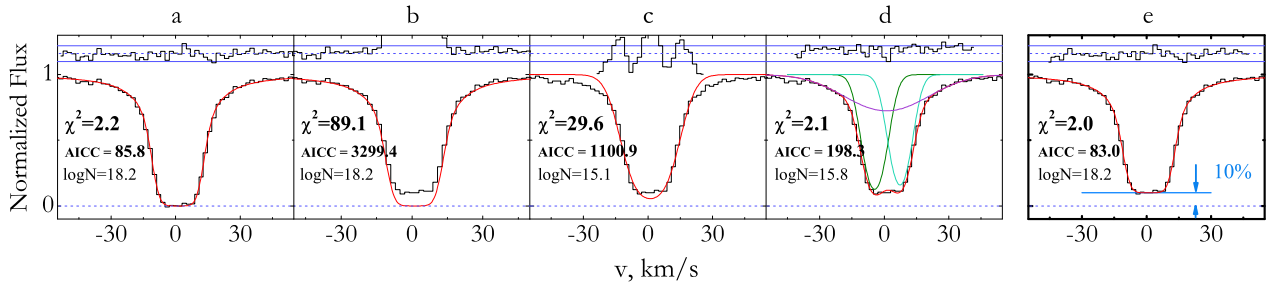}
        \caption{\rm Panel (a) shows a single-component H$_2$ absorption line with $\log N=18.2$
        and $b = 4$\,km\,s$^{-1}$ and a fit by a one-component model (red line). The flux at the bottom of the line
        goes to zero. We add a residual flux of about 10\,per cent (for clarity) to the data in panels (b), (c), (d)
        and (e). The flux at the bottom of the lines does not go to the zero level, and simultaneously the lines
        have Lorentzian wings. A simple one-component model cannot describe this line (cases (b) and (c)) and
        returns a large reduced $\chi^2$. However, if we add a new unresolved component to the model,
        the $\chi^2$ will be significantly decreased (case (d)) but not the statistical criterion, the AICC
        (see \citealt{King2011}), and the returned column density will be two order of magnitude too small.}
        \label{diff_profiles}
\end{center}
\end{figure*}

Partial covering is readily detectable in the spectrum of a background quasar if the cores of the
saturated absorption lines do not reach the zero flux level. This indicates that a part of the radiation
from the QSO passes by the cloud. The covering factor characterizing partial coverage is defined by the
ratio,
\begin{equation}
f_{\rm c}=\frac{F_{\rm cloud}}{F_{\rm total}},
\end{equation}
where $F_{\rm cloud}$ is the flux that passes through the absorbing gas and $F_{\rm total}$ is the total
flux. Therefore the measured flux in the spectrum, $F(\lambda)$, can be written as:
\begin{equation}
F(\lambda) = [F_{\rm total}(\lambda) - F_{\rm cloud}(\lambda)] + F_{\rm cloud}(\lambda)
\exp[-\tau(\lambda)] ,
\end{equation}
here $\,\tau(\lambda)\,$ is the optical depth of the cloud at the wavelength $\lambda$. The line flux
residual (LFR) is the fraction of the QSO flux which is not covered by the cloud. These definitions are
illustrated in Fig.\,\ref{Def_CF}. The determination of the covering factor is trivial in the case of
highly saturated absorption lines (see Fig.\,\ref{Def_CF} panels (a) and (b)) while for a partially
saturated line (see Fig.\,\ref{Def_CF} case (c)) the analysis requires a more sophisticated procedure. In
this case it is necessary to use several absorption lines originating from the same levels but with
different values of $\lambda f$, which is the product of the oscillator strength, $f$, and the wavelength
of the transition, $\lambda$. Such an analysis has been performed by \citet{Ivanchik2010} and more
precisely by \citet{Balashev2011} for the spectrum of Q\,1232$+$082, and  by \cite{Albornoz2014} for the
spectrum of Q\,0643$-$504. A similar situation was observed for HE0001$-$2340. \citet{Jones2010} have
considered the possibility of partial coverage of the BLR to explain the observed Mg\,{\sc ii} equivalent
widths. In contrast to the rare situations where partial covering occurs from intervening systems (see
also \citealt{Petitjean2000}), partial covering is typical for absorption systems associated with quasars
( e.g. \citealt{Petitjean1994,Rupke2005, Hamann2010, Muzahid2013}).

A failure to take into account the partial coverage effect in a spectroscopic analysis can lead to a
significant underestimation of the column density of an absorber. The systematic bias (of column density)
can exceed several orders of magnitude for saturated lines. As an example, consider an absorption line
which consists of one component and has the high column density. The spectrum and the corresponding
one-component model are shown in panel (a) of Fig.~\ref{diff_profiles}. In panels (b), (c), (d) and (e)
the same line is presented, but part of radiation from background source (10\,\% for clarity) passes by a
cloud. If we take into account the LFR, then the line can be properly fitted by a one component model and
we can recover the high input column density (with given accuracy) and measure the LFR value (panel (e) of
Fig.~\ref{diff_profiles}). If the residual flux is not taken into account, then the one component model --
Lorentzian (b) or Gaussian (c) profiles -- is not adequate, the reason being that a one-component Voigt
profile cannot describe the unsaturated bottom and far wings of the line simultaneously. Using additional
components, as shown in panel (d), a result with a satisfactory $\chi^2$ is obtained. However, this
solution is incorrect, because the resulting column density ($\log N = 15.8$) is much smaller than the
input one ($\log N = 18.2$). To distinguish between cases (d) and (e) we propose a new method based on an
analysis of several absorption lines with different oscillator strengths. The description of the method
and application to the analysis of H$_2$  in Q\,0528$-$250 are described in more detail later.

\section{Molecular hydrogen}
\label{H2analysis}

Molecular hydrogen lines are detected in the spectrum of Q\,0528$-$250 from the DLA at ${\rm
z_{abs}=2.811}$. Column density of neutral hydrogen in this DLA system is $\log N(\mbox{H\,{\sc i}}) =
21.35\pm0.07$ \citep{Noterdaeme2008}. H$_2$ lines correspond to transitions from rotational levels up to
${\rm J=5}$. To fit the molecular hydrogen lines, the spectrum has been normalized with a continuum
constructed by fitting the selected continuum regions devoid of any absorptions with spline.

\subsection{Number of components}
\label{numbcomp}

\begin{table}
\caption{Results of the previous analyses of the H$_2$ system at $z=2.811$ towards Q\,0528$-$250.} \label{refH2}
        \begin{center}
        \begin{tabular}{|l|c|c|c|c|}
        \hline
        \hline
        Year & $\log N_{\rm tot}$ & ${\rm N_{Comp.}}$ & Resolution & Ref.\\
        \hline
        1985 & 16.46$\pm$0.07  & 1 & & [1] \\
        1988 & 18.0            & 1 & 10 000 & [2]\\
        1998 & 16.77$\pm$0.09  & 1 & 10 000 & [3]\\
        2005 & 18.22$^{+0.13}_{-0.17}$  & 2   & 40 000 & [4]\\
        2006 & 18.45$\pm$0.02  & -- &  & [5] \\
        2011 & 16.56$\pm$0.02  & 3 & 45 000 & [6]\\
        {\bf 2015} & {\bf 18.28$\pm$0.02}& {\bf 2}  &  45 000  & {\bf This work} \\
        \hline
        \hline
        \multicolumn{5}{|l|}{[1] ~\citet{Levshakov1985}; [2]  ~\citet{Foltz1988};}\\
        \multicolumn{5}{|l|}{ [3] ~\citet{Srianand1998}; [4] ~\citet{Srianand2005};}\\
        \multicolumn{5}{|l|}{[5] ~\citet{Circovic2006}; [6]  ~\citet{King2011}}\\
        \end{tabular}
        \end{center}
        \label{results}
\end{table}

\begin{figure}
            \begin{center}
            \includegraphics[width=0.45\textwidth]{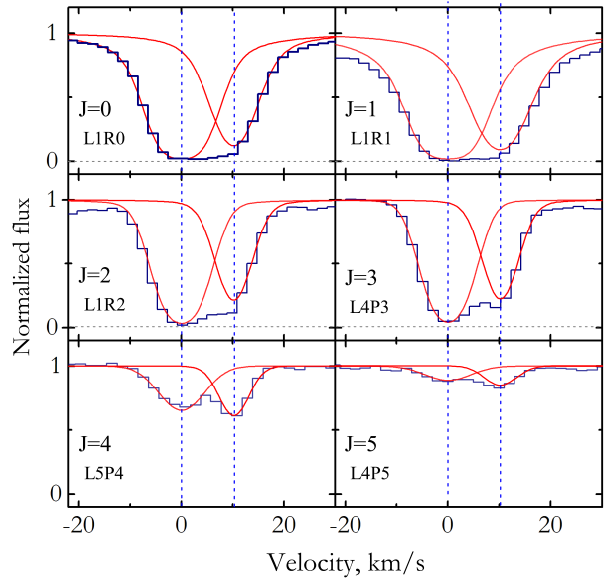}
            \caption{\rm Examples of H$_2$ line profiles corresponding to the transitions from ${\rm J=0-5}$ levels.
            Two components are clearly seen in the ${\rm J=4}$ and 5 lines. We have found that two components are
            enough to obtain a satisfying $\chi^2$. The origin of the velocity scale is taken at the redshift
            of the H$_2$ component, ${\rm z_A = 2.81099.}$}
            \label{H2_str}
            \end{center}
        \end{figure}

\begin{figure*}\begin{center}
       \includegraphics[width=0.97\textwidth]{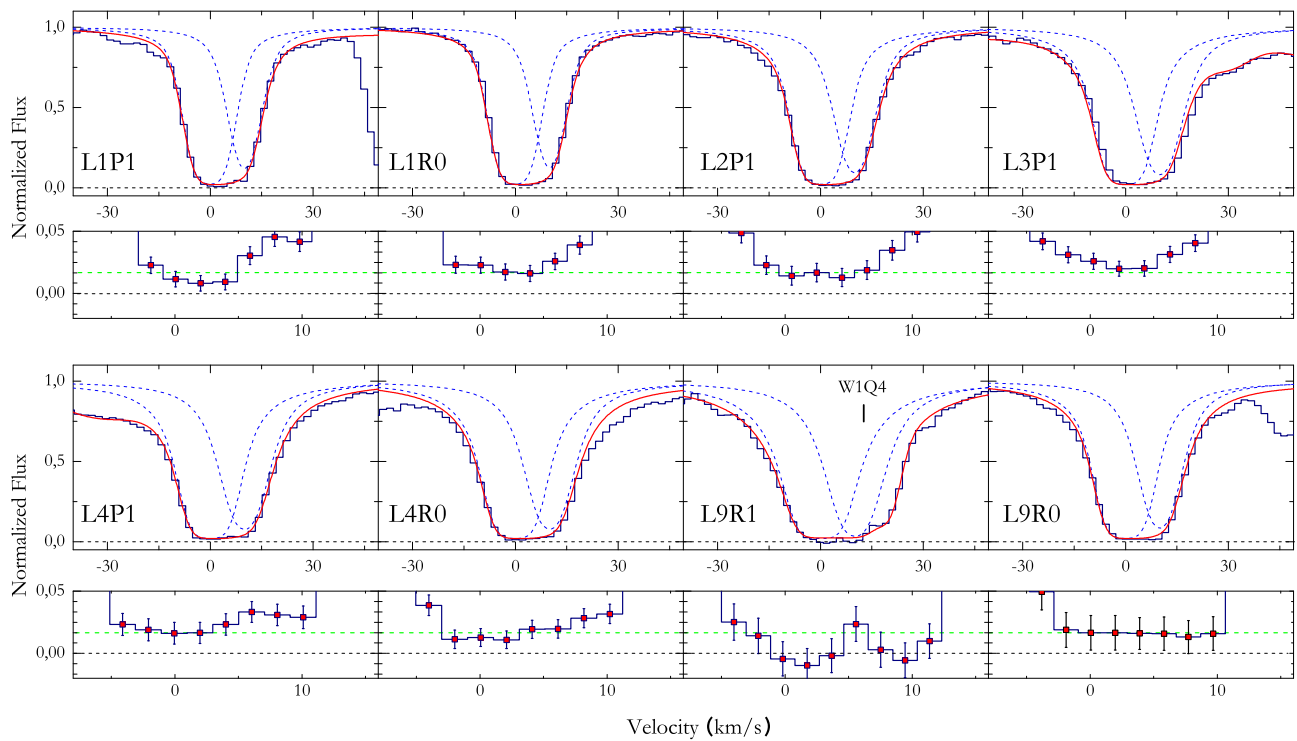}
        \caption{\rm Some of the absorption lines of H$_2$ from the ${\rm J=0,1}$ levels detected in the spectrum
        of Q\,0528$-$250 and the best Voigt profile fit (red line). Two components of the absorption system are shown
        by blue dashed lines. Only unblended saturated lines are present. The presence of prominent Lorentzian wings
        of the lines indicate high H$_2$ column density. The additional panels show a close-up of the bottom of
        the corresponding lines. The dashed black horizontal line represents the zero flux level. It is clearly seen
        that about 2\,per cent of the LFR is present at the bottom of the saturated H$_2$ lines. The x-axes show
        the velocity offset from the centre of H$_2$ component at ${\rm z = 2.81099}$.  In the profile of H$_2$
        line L9R1, some pixels in the bottom of the line are lower than 2 per cent level, which is probably because
        this line is blended with Ly$\alpha$ forest absorption lines.}
        \label{Lwing_ex}
\end{center}
\end{figure*}

The profiles of the H$_2$ absorption lines have a complex structure that cannot be fitted with a single
component.  At least two components are clearly seen in the lines corresponding to ${\rm J=4, J=5}$
rotational levels (see Fig.\ref{H2_str}). Since the first identification of the H$_2$ system in the quasar
\citep{Levshakov1985} other studies have been conducted, providing discordant results (see
Table~\ref{results}). \citet{Srianand2005} used a two-component model, while \citet{King2011} pointed out
that a three-component model is very strongly preferred over two-component model. \citet{King2011} used a
fitting procedure where they increased the number of components in the absorption system in order to
minimize the corrected Akaike information criterion, AICC\footnote{ This statistical criteria allows for a
choice of prefered model among several models with different numbers of fitting parameters. ${\rm
AICC}=\chi^2 + 2p + \frac{2p(p+1)}{n-p-1},$ where $p$ is the number of fitting parameters, $n$ is the
number of spectral points, included in an analysis.} \citep{Sugiura1978, King2011}. The resulting total
column density differs by two order of magnitude from \citet{Srianand2005}. In that case a criterion for
choosing preferred model is the consistent derived physical parameters of a cloud. It can be noted note
that by using H$_2$ column density reported by \citet{King2011} (based on the three-component model) a
$N{\rm{(HD)}}/2N{\rm{(H_2)}}$ ratio is obtained, that is about an order of magnitude higher than the
primordial one. Meanwhile, the H$_2$ column density in two-component model of \citet{Srianand2005} gives a
reasonable $N{\rm{(HD)}}/2N{\rm{(H_2)}}$ ratio, which is consistent with the typical values measured at
high redshift \citep{Balashev2010}.

A very important point for the choice of a reasonable absorption profile model in the case of
Q\,0528$-$250 is the presence of Lorentzian wings in ${\rm J=0}$, ${\rm J=1}$ line profiles (see
Fig.~\ref{Lwing_ex}). It is an indicator of the high H$_2$ column density of absorption system with $\log
N{\rm (H_2) > 18}$ which is consistent with the result reported by \citet{Srianand2005}. However, in the
new spectrum of Q\,0528$-$250 the H$_2$ lines with prominent Lorentzian wings have some residual flux at
the bottom which significantly differs from the zero flux level (see Fig.~\ref{Lwing_ex} and
Fig.~\ref{LymanForest}). As a consequence, the fit by \citet{Srianand2005} gives the large reduced
$\chi^2$. It is probable that the signal-to-noise ratio in the previous spectrum was insufficient to
detect the residual flux.

The current study shows that the residual flux detected in the bottom of saturated H$_2$ lines in the
spectrum of Q\,0528$-$250  is the result of a partial coverage effect. In this case, the profiles of H$_2$
lines can be very well fitted by two-component model. Therefore, there is no need to increase the number
of components in H$_2$ profiles (such as it is done by \citealt{King2011}) to explain the complex
structure of lines (an example is given in Fig.\,\ref{diff_profiles}). In the next three subsections, we
provide evidence of the existence of the partial coverage in the spectrum of Q\,0528$-$250.

\subsection{Zero-flux level correction}
\label{Zeroflux}

To measure the LFR in H$_2$ absorption lines we need to derive the zero flux level in the spectrum. A
non-zero flux in the core of a saturated absorption line can be the result of inaccurate determination in
between spectral orders of scattered light inside the instrument. The zero flux level in the spectrum can
be estimated using the saturated Ly$\alpha$ absorption lines which are numerous and almost uniformly
distributed over the wavelength range where H$_2$ absorption lines are located. Ly$\alpha$ lines are
associated with intergalactic clouds that are larger than several kpc, thus it is most likely that they
cover the background source completely.

\begin{figure*}
\begin{center}
        \includegraphics[width=0.97\textwidth]{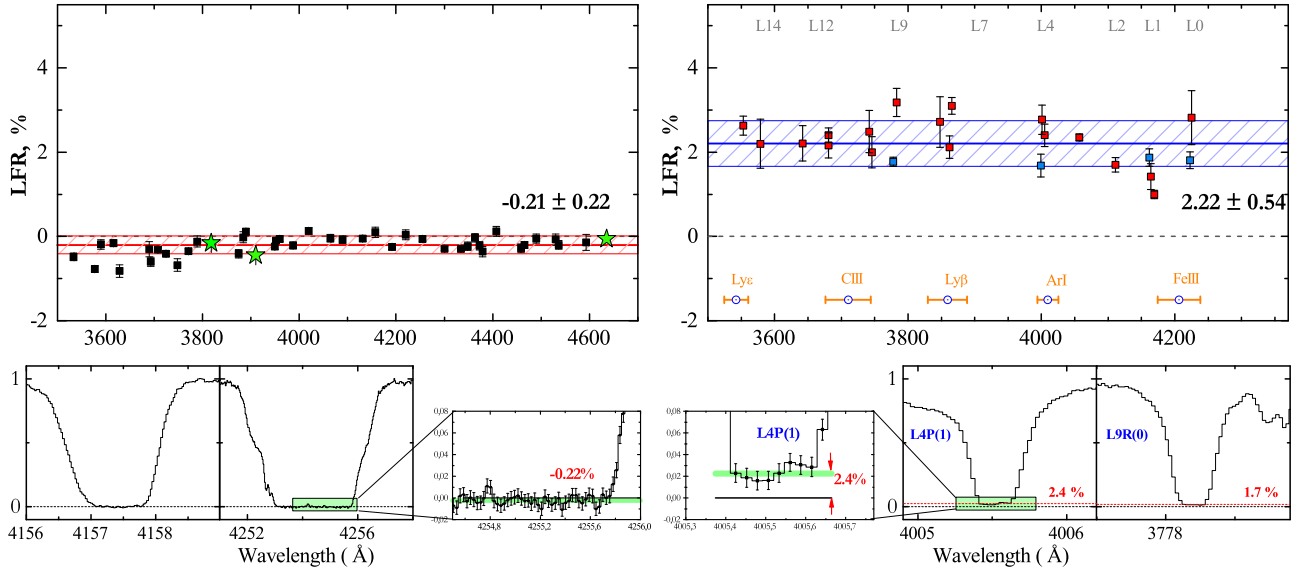}
        \caption{\rm Top left: Analysis of the zero flux level in the spectrum of Q\,0528$-$250.
        The LFRs measured at the bottom of the saturated absorption lines in the Ly$\alpha$  forest are plotted
        against the wavelength as filled squares. The residual fluxes are expressed in per cent of the continuum.
        The green stars show the LFR measured in the Ly$\beta$ and Ly$\alpha$ absorption lines of the DLA system
        at ${\rm z_{ abs}=2.79}$ and the Ly$\alpha$ absorption from the DLA system at ${\rm z_{abs}=2.14}$.
        The average value of the zero flux level is shown by the red horizontal line.  The scatter in the LFR at
        the bottom of the lines (i.e. noise in the spectrum) is found to be at the level of 0.22 per cent.
        Bottom left: a few examples of saturated Ly$\alpha$ lines. The third bottom panel is the zoom of a part
        of the second panel. Top right: the residual fluxes at the bottom of the H$_2$ absorption lines from
        rotational levels ${\rm J=0}$ (blue) and ${\rm  J=1}$ (red) are plotted against wavelengths as filled squares.
        The average value of the LFR in the H$_2$ lines is shown by the blue line. Blue circles and orange horizontal
        lines show the positions of QSO broad emission lines and their widths taken from \citet{VandenBerk2001}.
        Bottom right: A few saturated H$_2$ absorption lines are present. The red dashed lines show the estimated
        residual flux in each H$_2$ line. The black line indicates the zero level. The first panel shows the zoom of
        a part of the second panel. It can be seen that the flux in the pixels located at the bottom of saturated
        H$_2$ lines systematically departs from zero. The shift is about 2\,per cent of the continuum.}
\label{LymanForest}
\end{center}
\end{figure*}

\begin{figure}
            \begin{center}
            \includegraphics[width=0.45\textwidth]{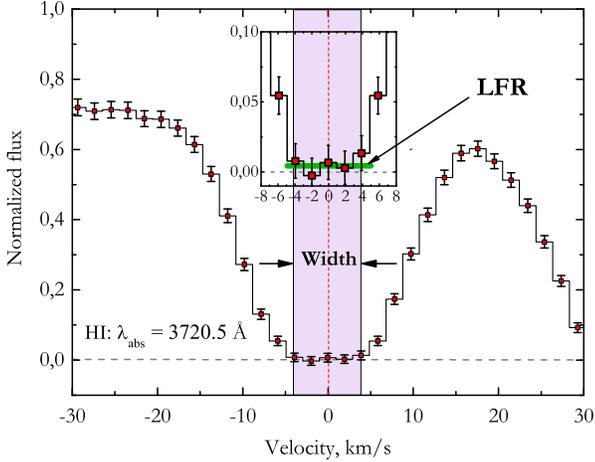}
            \caption{\rm An illustration of the residual flux measurement procedure.
            To estimate the residual flux we determine the median flux for pixels in a line profile
            that are in the range of $\pm1\sigma$ from the flux at the nearest pixel to the central wavelength
            $\lambda_c$. This region is shown by purple shading. We refer to the width of this region as the width
            of the bottom of the line.}
            \label{EQ_LFR}
            \end{center}
        \end{figure}

Wide flat bottom lines ($\Delta \lambda >1$ \AA, i.e. $\delta v > 80$~km\,s$^{-1}$) were selected, which
guarantees that lines are saturated and therefore measured fluxes in the bottom of the lines (LFRs) are
the real zero flux level in the spectrum. These lines were selected in the spectral region
3500$-$4700\,\AA. To estimate the residual flux at the bottom of the line the procedure illustrated in
Fig.\,\ref{EQ_LFR} was implemented. We selected several pixels in a line profile for which flux is within
$f_{c}\pm1\sigma_i$, where $f_{c}$ is the flux at the centre of the line and $\sigma_i$ is the error in
pixel $i$. The residual flux was calculated as the median of the flux in the selected pixels. The width of
the line bottom was estimated as the difference between the right most and left-most selected pixels from
the line centre (see Fig.\,\ref{EQ_LFR}). The top-left panel of Fig.~\ref{LymanForest} shows the LFR
values obtained for the selected saturated Ly$\alpha$ absorption lines. The average value is found to be
$(-0.21\pm0.04)\%$. The standard deviation of the points is $\sigma\simeq 0.22\%$. The green stars show
the LFR values measured at the bottom of the Ly$\alpha$ and Ly$\beta$ lines associated with the DLA-system
at ${\rm z_{abs}=2.79}$ and with the Ly$\alpha$ line of the second DLA-system at ${\rm z_{abs} = 2.14}$.
These lines are the most saturated lines for the spectrum.

\subsection{Partial coverage of H$_2$ absorption lines}
\label{Resflux}

\begin{figure*}
\begin{center}
        \includegraphics[width=0.95\textwidth]{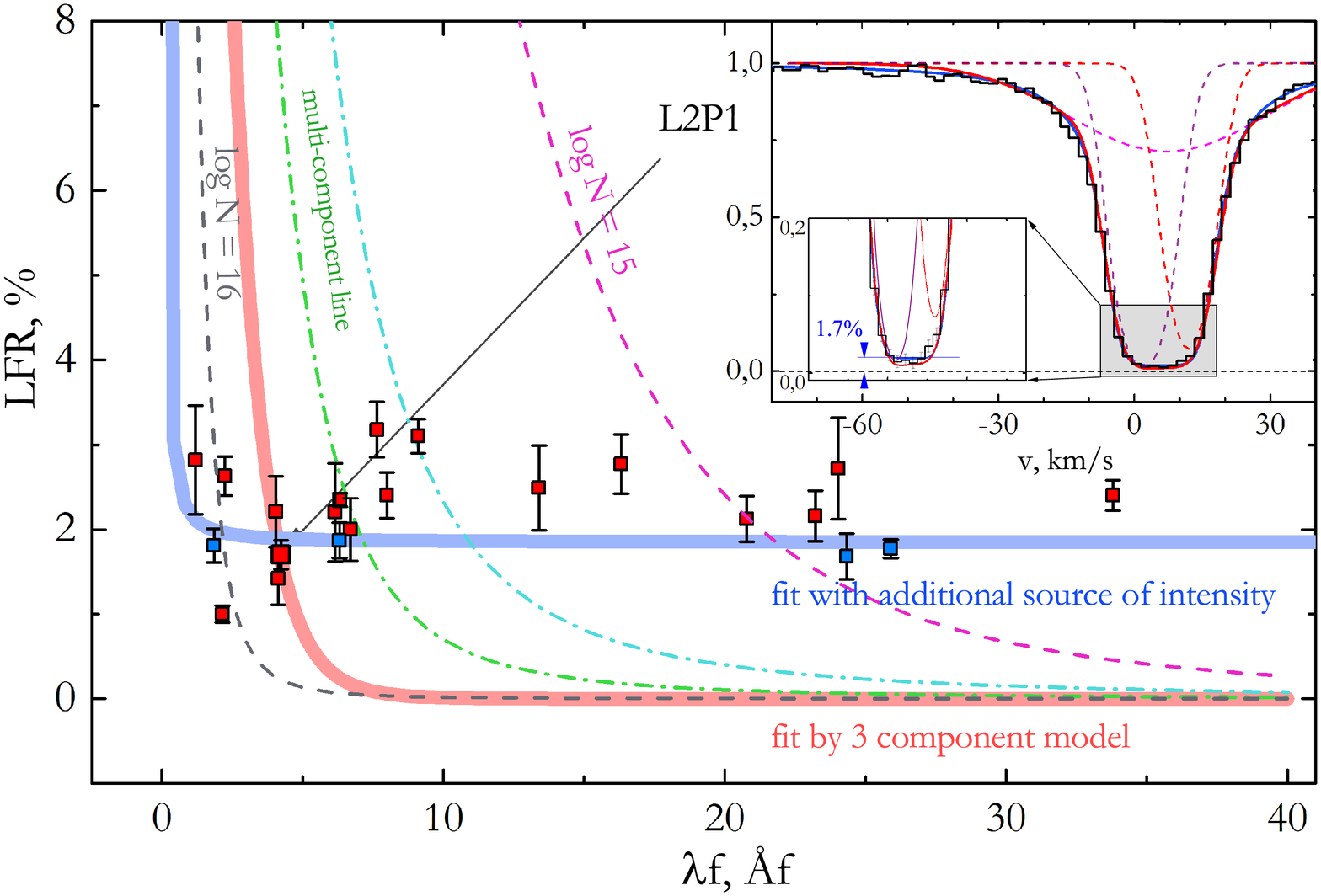}
        \caption{\rm The residual flux at the bottom of the H$_2$ absorption lines is given versus the parameter
        $\lambda f$ (where $f$ is the oscillator strength of a line). Blue and red points correspond to the H$_2$
        lines from the ${\rm J=0}$ and ${\rm J=1}$ levels. The theoretical flux at the bottom of the one-component line
        with a fixed column density $N$ and Doppler parameter $b$ is shown by the dashed curve.  The violet and grey
        curves correspond to column densities $\log N{\rm(H_2)}= 15$ and 16. It is seen that all points cannot be
        described by one curve simultaneously. The blue dash-dotted line represents the calculated residual flux at
        the bottom of a composite line, which consists of two components with $\log N{\rm(H_2)}= 15$ for both lines
        and a velocity separation of $\delta v = 2\,{\rm km\,s^{-1}}$. The green dash-dotted line presents
        the same model for three components with the same column densities and velocity separations of
        $\delta v = 2$ and 4 km\,s$^{-1}$, respectively (see text). An example is shown in the right-hand top panel.
        We fit a certain H$_2$ line (L2\,P1) by two models: one component with the LFR (blue line) and three components
        (red line). Both models give adequate fit of the line. But model without the LFR cannot describe flux
        in all lines together, as shown by the thick red line in the graph. In contrast, all points are described
        by a one-component model with residual flux (see the thick blue line).}
        \label{flambda}
\end{center}
\end{figure*}

To estimate the residual flux in H$_2$ lines we selected ${\rm J=0,1}$ lines without apparent blends. The
LFR was measured by the same technique as for the Ly$\alpha$ forest lines. The analysis found that the
flux in the bottom of the lines is quite constant over large velocity range $\delta v\simeq
10$\,km\,s$^{-1}$ (i.e. the dispersion of points is within the range of the average statistical error)
that is wider than the full width half maximum (FWHM) of the UVES (6\,km\,s$^{-1}$) and is comparable with
the shift between centres of the two components. Therefore, the structure of H$_2$ system has no effect on
the residual flux. The comparison of the obtained residual flux in the ${\rm J=0,1}$ H$_2$ lines with the
level of the residual flux in the Ly$\alpha$ forest lines is shown in Fig.\ref{LymanForest}. The filled
squares represent the residual flux in H$_2$ lines versus its location in the spectrum. We have estimated
the line flux residual at the level $2.22\pm0.54$\,\% of the continuum, which significantly exceeds the
zero-flux level.

Non-zero residual flux at the bottom of saturated lines can also be the result of the convolution of the
saturated lines with the instrumental function, or an imperfect data reduction, and/or of the blend of
several unresolved unsaturated lines. First, however H$_2$ lines are wide with widths larger than the FWHM
of the UVES spectrograph (6\,km/s,\,i.e.\,$\sim0.08$\,\AA), so after convolution with the instrument
function, the flux at the bottom of these lines must still go to zero. Secondly, the improper data
reduction is not a viable explanation of the residual fluxes of saturated Ly$\alpha$ lines are
consistently equal to zero. Also we tested the dependence of residual flux on the width of the bottom of
line for H$_2$ and for the Ly$\alpha$ forest lines. This is shown in Fig.\ref{AppA1}. Several lines in the
Ly$\alpha$ forest were detected; these have similar widths as the H$_2$ lines and go to the zero level.
The profiles of these lines are shown in Fig.\,\ref{AppA2}.

Lastly, the residual flux can not be a result of the composition of several unresolved unsaturated lines
because the lines have Lorentzian wings.

\subsubsection{The $\lambda f$ test}
An additional test which confirms the presence of the partial coverage effect is discussed here (and
illustrated in Fig.\,\ref{flambda}). Molecular hydrogen lines from ${\rm J = 0, 1}$ levels have very
different values of the product $\lambda f$ (from 0.6 for L0P1 to 36.0 for W1Q1). It is known that the
flux in the bottom of an absorption line decreases exponentially when $\lambda f$ increases, $F \propto
\exp(-\lambda f)$. The results of the calculation of this dependence for an absorption line in the
spectrum with VLT resolution (${\rm FWHM=6\,km\,s^{-1}}$) are shown in Fig.\,\ref{flambda} by the dashed
lines. For simplicity, the modeled line profile consists of one component. The Doppler parameter ${\rm
b=4\,km\,s^{-1}}$ and the Damping width $\Gamma_{lm}$ is the same as for the H$_2$ line L2P1. Because the
equivalent width of the line gets to the logarithmic part of the curve of growth, we do not consider the
different Doppler parameters. Two dashed curves (violet and grey) are calculated for the line with column
densities $\log N=15$ and 16. The curve is shifted from right to left as the column density $N$ increases.
For a higher column density, the residual flux equals zero for a wide range of $\lambda f$ (because the
line is saturated) and differs from zero only for small  values of $\lambda f \le 5$ (the case of an
unsaturated line). In a case of a multicomponent line (i.e. the blend of two or three lines with small
column density $\log N=15$), the residual flux in the bottom of the blended line would also behave
exponentially (dashed-dot curves). However, the behavior of the residual fluxes in the bottom of the
saturated H$_2$ lines in the spectrum of Q\,0528$-$250 is quite different. The residual fluxes in the
H$_2$ lines are shown by the filled blue and red squares. The squares do not follow the expected behavior,
moreover the points scatter similarly around a median value in the whole range of $\lambda f$.

Two models were applied to obtain a consistent and correct fit to the H$_2$ lines. Model (i) considers the
best fit of the certain H$_2$ line (e.g. L2P1), which has a value of $\lambda f$  = 4.23 and ${\rm LFR =
1.7\,\%}$ of the total flux. The line has a wide flat bottom $(\delta v\sim10\,{\rm km\,s^{-1}})$ and
Lorentzian wings. To fit this line without the partial coverage, it is necessary to use several
unsaturated components in the line profile. Therefore, this model describes the line profile well. Model
(ii) takes into account the partial coverage, and the line L2P1 can be fitted by the one-component model
with high column density $\log{\rm \,N=18.2}$ and ${\rm LFR = 1.7\,\%}$ (the blue line in the right-hand
top panel). This model also describes the line profile accurately. Using only one H$_2$ line we cannot
determine the most probable model. However, if we consider several H$_2$ lines from the same J level which
cover a wide range of $\lambda f$, we will be able to discriminate it. It is known that the lines from the
same J level correspond to the same physical region of a molecular cloud; therefore the lines are
described by the same set of physical parameters (the column density ${\rm N}$ and Doppler parameter ${\rm
b}$). The difference between line profiles originating from the same J level is caused only by the
different values of $\lambda f$ for the lines. Therefore, the correct model of an absorption system must
describe all measurements of residual flux in the bottom of lines from a given J level simultaneously.
Using the best fit parameters for models (i) and (ii) we have calculated the dependence of the residual
flux on $\lambda f$ (thick red and blue curves, respectively). The thick red line cannot describe all
squares in the main panel simultaneously, whereas the blue thick line can.

\subsection{Voigt profile Fitting}
\label{VpF}

A Voigt profile fitting of the H$_2$ absorption lines was performed, taking into account the partial
coverage. To describe a complex structure of line profiles we have divided the total flux detected by an
observer into two parts: from a main source and an additional one. In addition, we consider the H$_2$
system to be composed of two components (A and B) at redshifts ${\rm z_{A} = 2.81099}$ and ${\rm z_{B} =
2.81112}$. The light from the main source ($\sim98\%$ of the total flux) is intercepted by the two H$_2$
components and does not produce any residual flux in the saturated lines. The light from the additional
source that passes by H$_2$ clouds is not absorbed and therefore produces a uniform residual flux in H$_2$
lines. The flux in the absorption line of two components A and B can be described as
\begin{equation}
    \label{eq01}
    F(\lambda) = (1-f_c) F_{total} + f_c  F_{total}e^{-\tau_{\rm A}(\lambda)} e^{-\tau_{\rm B}(\lambda)},
\end{equation}
where $f_c$ (in relative units) is the covering factor for H$_2$ lines.

However, because the physical conditions in the A and B clouds (such us linear size, volume density, etc.)
might be different, we can expect the covering factors of quasar emission regions by two H$_2$ clouds also
to be different. In this case the construction of the H$_2$ line profiles is more complicated and we
present an analysis of this case below (see Appendix\,\ref{model2}). Here, it is important to note that
taking into account two covering factors does not allow for a better fit to the H$_2$ lines (see the
discussion in Appendix\,\ref{model2}).

Then the absorption lines for each $J$ level were described using seven fitting parameters: z$_{\rm A}$,
z$_{\rm B}$, b$_{\rm A}$, b$_{\rm B}$, N$_{\rm A}$, N$_{\rm B}$, $f_c$. We used uniform values of $f_c$
over the whole wavelength range, because the residual flux in the H$_2$ lines is almost independent of the
wavelength (see Fig.\,\ref{LymanForest}). The Doppler parameter b is a function of the rotational level
$J$. To estimate fitting parameters Markov Chain Monte Carlo (MCMC) method was implemented, and to speed
the convergence the Affine invariant ensemble sampler by \citet{Goodman2010} was used. The main advantage
of this searching algorithm is to better explore a parameter space and to avoid using the partial
derivatives of the $\chi^2$ function which eases a number of numerical issues. This allowed for more
reliable estimates of the fitting parameters in comparison with other algorithms.

\subsection{Fitting results}

The H$_2$ absorption system at $z=2.811$ has more than 130 absorption lines from ${\rm J=0}$ to ${\rm
J=5}$ rotational levels. For the current analysis, we selected H$_2$ lines that are free of any obvious
blends. The sample examined contains 99 lines. The best fit of H$_2$ lines are shown in
Fig.\ref{J0lines}$-$Fig.\ref{J5lines} ranked following wavelength positions. The best-fitting parameters
are illustrated in Table~\ref{table_res2}. The reduced $\chi^2$ is 1.08 (the number of fitting points
$\sim 1500$).

The H$_2$ absorption system is highly saturated. The total H$_2$ column densities are $18.10\pm0.02$  and
$17.82\pm0.02$ for the A and B components consequently. This is consistent with the presence of the
Lorentzian wings in the profiles of ${\rm J=0}$ and ${\rm J=1}$ levels. The obtained orto--to--para ratios
are $2.7\pm0.1$ and $3.2\pm0.3$ for the A and B components, respectively. The corresponding kinetic
temperatures of the H$_2$ clouds are ${\rm T_{01,A}=141\pm6\,K}$ and ${\rm T_{01, B} = 167\pm13\,K}$.

\begin{figure*}
            \begin{center}
            \includegraphics[width=0.97\textwidth]{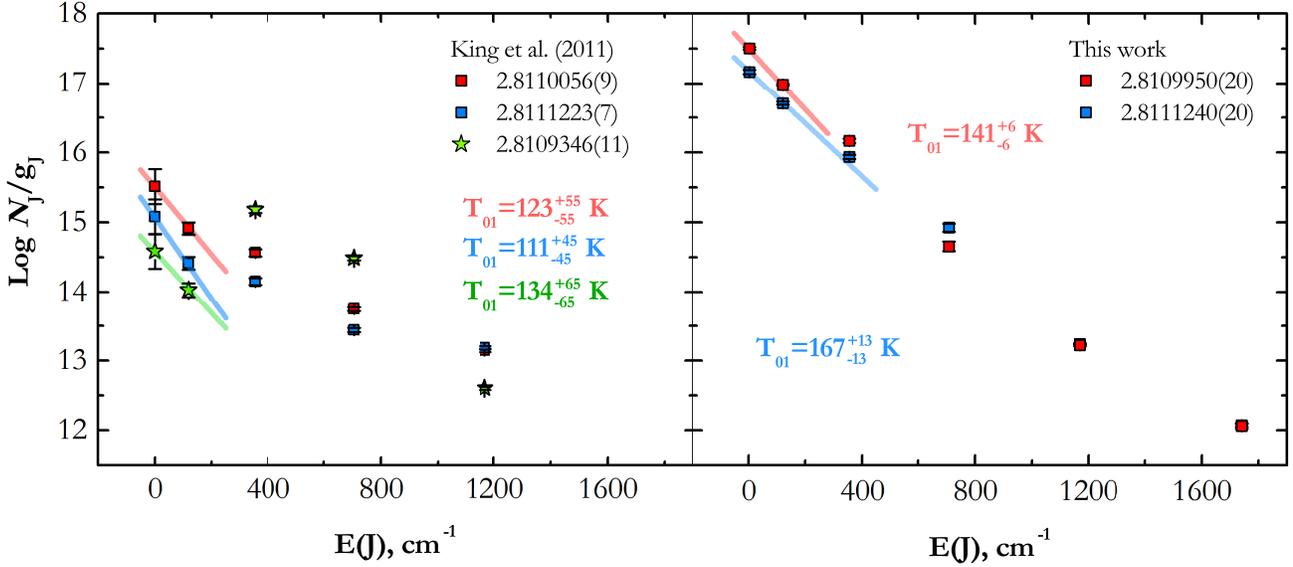}
            \caption{\rm The excitation diagram for H$_2$ towards Q\,0528$-$250. Here, $N_J$ is the column density of the transition from the $J$ level with $g_J$ degeneracy. Solid lines (blue, red and green) correspond to the excitation temperature ${\rm T_{01}}$ derived from $J=0$ and $J=1$ levels. The left-hand panel shows the result of the analysis performed by \citet{King2011}. The excitation diagrams for different components of the H$_2$ system are shown by blue, red and green colors. The right-hand panel shows the results from our present work. Instead of three unsaturated components we use two components with higher H$_2$ column densities.}
            \label{exc}
            \end{center}
\end{figure*}

Fig.\,\ref{exc} shows a comparison of the H$_2$ excitation diagrams for our measurements and those of
previous work. The left-hand panel shows the result of the analysis performed by \citet{King2011}, the
right-hand panel shows the result from the present work. Because the results presented here are close to
data reported by \citet{Srianand2005}, these are not shown here. We show excitation diagrams for two H$_2$
components at ${\rm z_{A}=2.81099}$ (2.811001 found by \citet{King2011}) and ${\rm z_{B}=2.81112}$. The
third H$_2$ component found by \citet{King2011} at ${\rm z_{C}=2.8109346(11)}$ is represented by the green
stars. Although the ratio of ${\rm J=0}$ and ${\rm J=1}$ levels of third component is the same as in other
components, the excitation diagram is not physically realistic (the excitation temperatures ${\rm T_{02}}$
and ${\rm T_{13}}$ for the third component are negative). This might be the result of the incorrect model
being used. It is seen, that the discrepancy between our results and those of \citet{King2011} is larger
only for the low J levels, where the influence of the partial coverage effect on the structure of the line
profiles is significant. For high J levels, where column densities of H$_2$ are less, the results agree in
$1\sigma$. To sum up, the total column density of H$_2$ from the measurement is 18.284\,$\pm$\,0.025, that
is about two orders of magnitude larger than the value reported by \citet{King2011} 16.556\,$\pm$\,0.024.

It should be noted, that the same values of $N$ and b parameters for all transitions of one J level were
used. Using the model  with taking into account the partial coverage effect we obtain the reduced
$\chi^2\simeq1$ without an increase of the statistical errors of the spectrum. This is important, because
in the previous analysis of H$_2$ system in Q\,0528$-$250 \cite{King2011} noted that without artificially
increasing the statistical errors, the reduced $\chi^2$ was $\gg1$ (see the caption of table\,6 in
\citealt{King2011}).

The value of the residual flux in H$_2$ lines is fitted as an independent parameter of an analysis. The
best value is $2.40\pm 0.07$\,\% of the continuum, which agrees with the average value of the residual
flux obtained from the analysis of ${\rm J=0,1}$ H$_2$ lines (see Section\,\ref{Resflux}).

\begin{table}
\caption{Column densities and Doppler parameters obtained from Voigt profile fitting of the H$_2$ system at ${\rm z_{abs}=2.811}$ towards Q\,0528$-$250 after taking care of partial coverage.} \label{table_res2}
\begin{tabular}{|c|c|c|c|c|} 
\hline \hline
\!System\! & J & ${\rm z_{abs}}$ & $\log N\,{\rm (cm^{-2}})$ & b~(km s$^{-1}$)\\
\hline
\hline
A  &0 & ~~2.8109950(20)  & 17.50 $\pm$ 0.02 & ~2.66 $\pm$ 0.05 \\
   &1 & ~~2.8109950(20)  &  17.93 $\pm$ 0.01 & ~2.71 $\pm$ 0.05 \\
   &2 & ~2.8109952(5) &  16.87 $\pm$ 0.03 & ~2.75 $\pm$ 0.03 \\
   &3 & ~2.8109934(5) &  15.97 $\pm$ 0.07 & ~2.87 $\pm$ 0.07 \\
   &4 & ~2.8109938(8) &  14.18 $\pm$ 0.01 & ~4.79 $\pm$ 0.11 \\
   &5 & ~2.8109938(8) &  13.58 $\pm$ 0.02 & ~5.04 $\pm$ 0.39 \\
&&&&\\
B &0 & ~~2.8111240(20)  & 17.16 $\pm$ 0.03 & ~1.17 $\pm$ 0.06 \\
&1    & ~~2.8111230(20)  & 17.67 $\pm$ 0.02 & ~1.14 $\pm$ 0.06 \\
&2    & ~2.8111235(7) & 16.64 $\pm$ 0.03 & ~1.22 $\pm$ 0.02 \\
&3    & ~2.8111238(6) & 16.24 $\pm$ 0.06 & ~1.25 $\pm$ 0.03 \\
&4    & ~2.8111231(6) & 14.20 $\pm$ 0.01 & ~1.72 $\pm$ 0.09 \\
&5    & ~2.8111231(6) & 13.60 $\pm$ 0.02 & ~2.38 $\pm$ 0.45 \\
\hline
\hline
\end{tabular}
\end{table}

\section{HD molecules}
\label{HD}

\begin{figure}
      \begin{center}
              \includegraphics[width=0.47\textwidth]{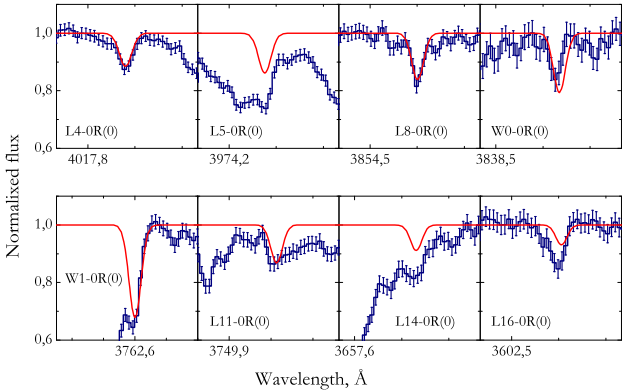}
              \caption{\rm The line profiles of HD molecules in the absorption system at ${\rm z_{abs}=2.811}$ in the spectrum of Q\,0528$-$250. HD lines are detected in component B only. The fit to the W0-0R(0) line is inconsistent owing to a poor definition of the local continuum near the base of the Ly$\alpha$ line of the second DLA system at ${\rm z_{abs}=2.14}$.}
              \label{HDprofiles}
      \end{center}
\end{figure}

\begin{figure}
      \begin{center}
              \includegraphics[width=0.47\textwidth]{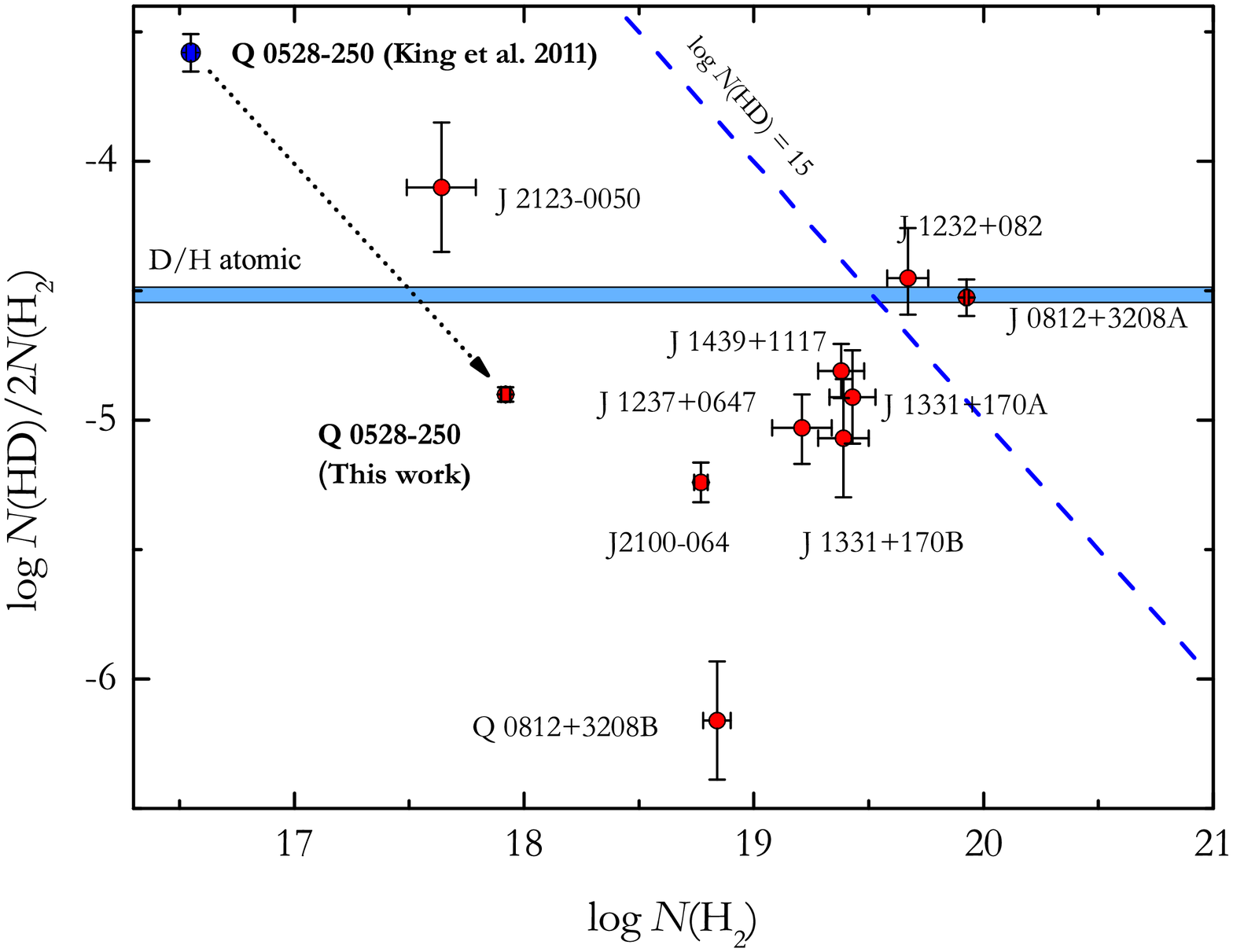}
              \caption{\rm The measurements of $N({\rm HD})/N({\rm H_2})$ vs $N({\rm H_2})$ in absorption systems at high redshift. The data are taken from \citet{Ivanchik2015}. The estimates of $N({\rm HD})/N({\rm H_2})$ toward Q\,0528$-$250 from this work and \citet{King2011} are shown by red and blue circles, accordingly. The blue horizontal strip represents the ratio of atomic D\,{\sc i} and H\,{\sc i} measured in quasar spectra \citep{Olive2012}. The solid dashed line corresponds to constant column density of HD, $\log N({\rm HD})=15$.}
              \label{HDtoH2}
      \end{center}
\end{figure}

In the new spectrum (obtained by co-adding all previous and additional new observations) the molecular HD
lines  associated with the H$_2$ absorption system were detected (using improved laboratory wavelengths,
\citealt{Ivanov2010}). The presence of HD lines in this system was first reported by \citet{King2011}. The
HD molecular lines are present only in the component B. Some of the HD lines are shown in
Fig.\,\ref{HDprofiles}. We have estimated the HD column density $\log{ N\rm(HD)_{\rm B}}=13.33\pm0.02$ by
analysis of the two most prominent unblended absorption lines, L4-0R(0) and L8-0R(0). The results of Voigt
profile fitting are presented in Table\,\ref{table_HD}. Other HD lines are highly blended and cannot be
used in the analysis. The obtained column density is significantly less than that required to produce
self-shielding, $\log{N\rm(HD)}\simeq\,15$, thus we can set only a lower limit to the isotopic ratio D/H
in the cloud. Because the total column density of H$_2$ in the component B is $\log
N{\rm(H_2)}=17.85\pm0.02$, we estimate ${\rm{D}}/{\rm{H}}\ge
N{\rm{(HD)}}/2N{\rm{(H_2)}}=\left(1.48\pm0.10\right)\times10^{-5}$. The obtained HD column density is
close to the result, $\log N{\rm(HD)} = 13.267 \pm 0.072$, reported by \citet{King2011}. However, taking
into account the significantly larger H$_2$ column density of the component B we have obtained a lower
value of $N{\rm(HD)}/2N{\rm(H_2)}$ than the result by \citet{King2011}. This limit is consistent with D/H
ratio obtained from the analyses of atomic species in quasar spectra (e.g. \citealt{Olive2012}). The
comparison of this result with other $N{\rm(HD)}/2N{\rm(H_2)}$ measurements at high redshift is shown in
Fig.\,\ref{HDtoH2}. The $N{\rm(HD)}/2N{\rm(H_2)}$ in this system is consistent with other values and
correspond to predictions of deuterium chemistry models of diffuse ISM clouds (e.g. \citealt{Balashev2010,
Liszt2014}).

Note that in the component B we detect  HD and C\,{\sc i} whereas in the component A these species are not
present, despite the higher H$_2$ column density of this component in comparison with the component B. As
for the component A we set an upper limit for HD and C\,{\sc i} column densities which are $\log{
N\rm(HD)_{\rm A}}\le13.1$ and $\log N({\rm C\,\mbox{\sc i}})_{\rm A} \le 12.0$. The lack of HD and C\,{\sc
i} in the component A might be the result of higher local UV radiation (e.g. bright young stars near cloud
A). It can destroy HD and ionize C\,{\sc i} while the H$_2$ molecules self-shield against the local
stellar radiation because to their high column density.

\begin{table}
    \caption{Best fitting parameters for HD molecular lines in the spectrum of Q\,0528$-$250.} \label{table_HD}
    \begin{tabular}{|c|c|c|}
            \hline \hline
                    ${\rm z_{abs}}$ &  ~~~~$\log N\, {\rm (cm^{-2})}$~~~~ & ${\rm b\, (km\,s^{-1})}$\\
            \hline
                 ~2.811121\,$\pm$\,0.000002  & ~13.33\,$\pm$\,0.02 & ~2.25 $\pm$ 0.53  \\
            \hline
                    ~{HD/2H$_2$}& ~~(1.48\,$\pm$\,0.10)\,$\times\,10^{-5}$ & ~ \\
            \hline
            \hline
    \end{tabular}
\end{table}

\section{Discussion}
\label{discussion}
\noindent

The presence of a residual flux at the bottom of the saturated lines of the H$_2$ system at ${\rm
z_{abs}=2.811}$ towards Q\,0528$-$250 can be interpreted according to the arguments described in the
following subsections.

    \subsection{Unresolved multicomponent quasars}

At redshift ${\rm z\sim2}$, the internal structures in the emitting regions of of quasars with transverse
dimensions $\lesssim 3\, {\rm kpc}$ are unresolved by UVES observations. For example, binary quasars with
separations of the order of ${\rm \sim10\,kpc}$ \citep{Hennawi2006,Vivek2009} might remain unresolved.
Partial coverage can arise if Q\,0528$-$250 has a complex multicomponent structure and if not all of the
components are covered by the absorbing H$_2$ clouds.

The available Very Long Baseline Array (VLBA) image of Q\,0528$-$250 \citep{Kanekar2009, Srianand2012}
shows an unresolved component containing $\sim94$ per cent  of the total flux in the radio band (see
table\,6 of \citealt{Srianand2012}). Another $\sim6$ per cent is probably emitted by a diffuse component.
However, it is intriguing to note the consistency of these numbers with our findings. The spatial size of
radio emission core component is $65\times380$\,pc \citep{Kanekar2009}, that is significantly larger than
the size of the H$_2$ clouds. We have also looked at the images of PKS\,0528$-$250 in 13 and 4\,cm taken
as part of National Radio Astronomy Observatories' VLBA calibrators. The sources are unresolved even at
higher resolution achieved in these images.

    \subsection{Dust scattering}

Scattering by dust is characterized by a narrow radiation pattern. About 90\,per cent of the total flux is
scattered towards the observer within less than 5-deg opening angle (e.g. \citealt{Drain2003}). The size
of the scattering region can be much larger than the size of an H$_2$ cloud. This is why, even in the case
of total coverage of the QSO by an H$_2$ cloud, the scattered radiation can be registered as a residual
flux in the bottom of H$_2$ lines. The scattered flux by a dust-rich region of a DLA system can be
estimated as
\begin{equation}
    F_{\rm sc} = F \left(1-e^{-\tau_{\rm DLA}}\right) \frac{\Omega_{\rm QSO}^{*}}{\Omega_{\rm QSO}} \frac{\Omega_{\rm DLA}}{\Omega_{\rm sc}},
\end{equation}
where $F$ is the flux that passes through an H$_2$ cloud and is registered by the observer, $\tau_{\rm
DLA}$ is the mean optical depth along the line of sight toward the quasar, $\Omega_{\rm QSO}$ and
$\Omega_{\rm DLA}$ are the solid angles (measured by an observer on Earth) of the QSO emission region and
the scattering region, accordingly. $\Omega_{\rm QSO}^{*}$ is the solid angle of the QSO emission region
measured by an observer at the position of the DLA and $\Omega_{\rm sc}$ is the solid angle in which most
of radiation is scattered by dust. Therefore, the residual flux in the bottom of H$_2$ lines is $LFR =
F_{sc}/F$. Assuming that a typical dimension of the scattering region is as large as $1-10\,{\rm kpc}$,
LFR can be of the order of several percent. This is consistent with observations. The scattered radiation
can produce a non-zero residual flux for metal lines also. However, this is not detected for
Q\,0528$-$250.

    \subsection{Jet-cloud interaction}

The jet emission pointing toward us contributes to the continuum in the optical band. The extended part of
the jet interacts with the external ISM and can warm up a cloud which is distant from the central source.
Similar situation has been detected in the deep radio-optical-X-ray observations of the radio galaxy
PKS\,B2152$-$699 by \citep{Worrall2012}, who have reported the first high-resolution observations of the
radio jet in the direction of an optical emission-line high ionization cloud. The {\it Hubble Space
Telescope} image shows not only emission in the region of the high-ionized cloud -- interpreted as ionized
gas by \citet{Tadhunter1988} and \citet{Fosbury1998} -- but also emission associated with a radio knot.
The measured optical flux density of this knot at $4.97\,\times\,10^{14}$\,Hz is 2.5\,$\pm$\,0.2 $\mu$Jy
\citep{Worrall2012}.
This is strikingly similar to the value of the residual flux in the  V optical band which is ${\rm
2.4\pm0.1\,\mu Jy}$ (using the apparent magnitude ${m_{\rm V}=17.34}$ of Q\,0528$-$250, \cite{Veron2010}).
The variability of a jet emission (during from a few days up to a month) should be high and should also
induce variability of the residual flux in the H$_2$ line. This could be tested by further observations.

\section{Conclusion}
\label{conclusion}
\noindent

We have performed a detailed analysis of the H$_2$ system at ${\rm z_{abs}=2.811}$ towards the quasar
Q\,0528$-$250. This is a well-known system and the first detection of H$_2$ molecules at high redshift
\citep{Levshakov1985}. The system has been analysed in a variety of research, but the results obtained are
not consistent (see Table \,\ref{results}). The discrepancies can be explained by the presence of a
residual flux in the bottom of H$_2$ lines, which has not been considered in previous research. We have
derived the mean value of the residual flux as $(2.22\pm0.54)$ per cent of the continuum. This is
significantly higher than the zero flux level, $(-0.21\pm0.22)$ per cent, determined by analysis of the
Ly$\alpha$ forest lines.

Taking into account the residual flux in the H$_2$ lines we have obtained a consistent fit of the H$_2$
system using a two-component model with high column densities. The derived total column densities of
components A and B are $\log N{\rm(H_2)=18.10\pm0.02}$ and $17.82\pm0.02$, respectively.

We have performed the analysis of HD absorption lines detected only in component B. The estimated column
density is $\log N{\rm (HD)}= 13.33 \pm 0.02$ and thus we derive
$N{\rm{(HD)}}/2N{\rm{(H_2)}}=(1.48\,\pm0.13)\times10^{-5}$. This value is consistent with other
measurements of $N{\rm{(HD)}}/2N{\rm{(H_2)}}$ in quasar spectra at high redshift and can be considered as
a lower limit of the primordial deuterium abundance (Balashev et al. 2010).

Some interpretations for the presence of the residual flux are being offered here: (i) a multicomponent
quasar; (ii) scattering by dust; (iii) a jet-cloud interaction. We favour the latter interpretation (iii).
However, new optical and radio observations of Q\,0528$-$250 are necessary to confirm this and reject the
others.

We argue that taking into account partial coverage effects is crucial for any analysis of H$_2$ bearing
absorption systems in particular when studying the physical state of high-redshift ISM.

\vspace{2mm}{\footnotesize {\rm Acknowledgments.} } This work is based on observations carried out at the
European Southern Observatory (ESO) under programmes ID 66.A-0594(A) (PI: Molaro), ID 68.A-0600(A) (PI:
Ledoux), ID 68.A-0106(A) (PI: Petitjean) and ID 082.A-0087(A) (PI: Ubachs) with the UVES spectrograph
installed at the Kueyen UT2 on Cerro Paranal, Chile.

 The work is supported by Dynasty foundation and by
the RF President Programme(grant MK-4861.2013.2). RS and PPJ gratefully acknowledge support from the
Indo-French Centre for the Promotion of Advanced Research (IFCPAR) under Project N.4304-2.

\bibliographystyle{mn2e}
\bibliography{Q0528}

\appendix
\section{ Test of a model with two additional sources of residual flux}
\label{model2} 
\begin{figure}
            \begin{center}
            \includegraphics[width=0.48\textwidth]{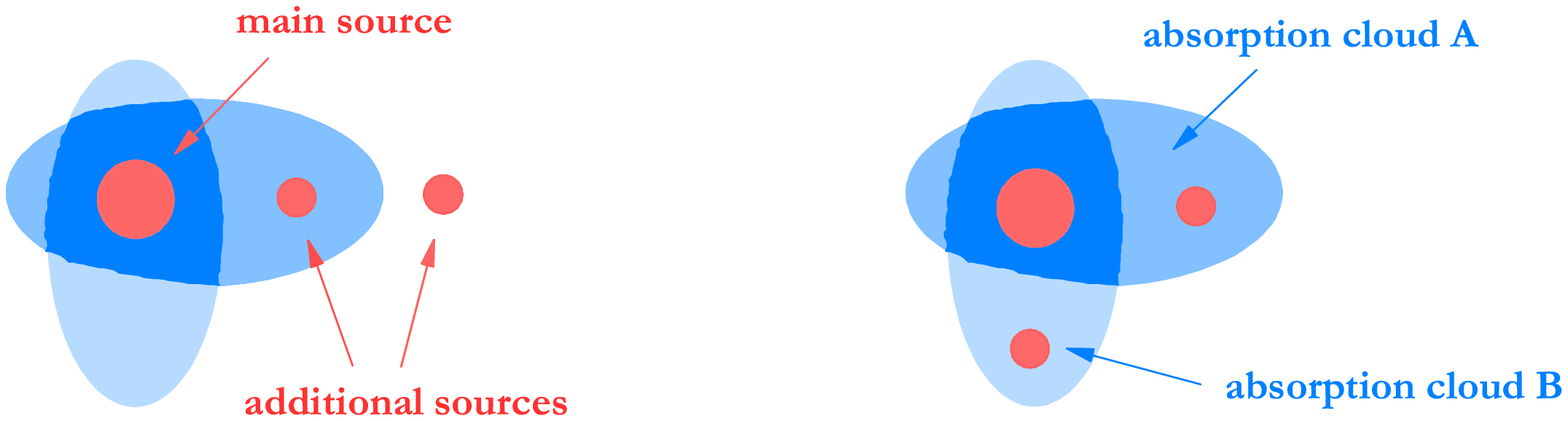}
            \caption{\rm An illustration of the possible explanation of the partial coverage effect for a case with independent sources of radiation (see text for detail).}
            \label{AppB1}
            \end{center}
        \end{figure}

{In this section, we consider a model of absorption system, where the projected area over
the illuminating source is different for components A and B. In that case a structure of H$_2$ line
profiles is more complicated. To describe a flux detected by an observer, we divide a quasar emission
region into three parts: the main source and two additional sources. The main source is covered by both
H$_2$ clouds and does not produce the residual flux in absorption lines. Additional sources can produce
the residual flux in two different ways (depending on the geometry of additional sources and H$_2$ clouds,
see Fig.\,\ref{AppB1}) as follows.

(i) One source (not covered by both clouds) produces the same residual flux in profiles of both H$_2$
systems. Another source is covered by one cloud and produces residual flux only in profile of the second
system:
\begin{equation}
\begin{split}
    \label{eq01}
    F(\lambda) = &m F_{total}(\lambda) + n F_{total}(\lambda)e^{-\tau_{\rm A}(\lambda)} + {}\\
    &+  (1 - m - n) F_{total}(\lambda) e^{-\tau_{\rm A}(\lambda)-\tau_{\rm B}(\lambda)}.
\end{split}
\end{equation}

(ii)    Each source produces the residual flux only for one system:
   \begin{equation}
    \begin{split}
        \label{eq01}
    F(\lambda) = &m F_{total}(\lambda)e^{-\tau_{\rm B}(\lambda)} + n F_{total}(\lambda)e^{-\tau_{\rm A}(\lambda)} +{}\\
    & +(1 - m - n) F_{total}(\lambda) e^{-\tau_{\rm A}(\lambda)-\tau_{\rm B}(\lambda)},
    \end{split}
    \end{equation}
where {\it m} and {\it n} indicate  intensities of additional sources in relative units. The results of
Voigt profile fitting of H$_2$ lines using different models of the H$_2$ system are presented in
Table\,\ref{table6}. The values of AICC for models {\it A} and {\it B} are significantly lower than for
model {\it C}, which could indicate the presence of one additional source of quasar radiation which
uncovered by both H$_2$ clouds. Taking into account the second additional source (for system B) does not
dramaticaly changed the AICC value and we have not found strong evidence for choosing a preferred model.
Therefore, we use the simplest model (with one additional source) to analyse the H$_2$ system.}

\begin{table}
\caption{\label{table6} A Comparison between models of the H$_2$ system with a different number of
additional sources of radiation. {\it A} indicates the model with one additional source (which used in
this work). {\it B} and {\it C} correspond to models (i) and (ii) described in see App.\,\ref{model2}. $p$
given the number of fitting parameters, AICC is the Akaike information criterion (see
Sect.\,\ref{numbcomp}). $m$ and $n$ indicate the intensity of  additional sources relative to the total
flux of a quasar.}
    \begin{tabular}{|c|c|c|c|c|c|}
            \hline
            Model &  $p$ & AICC & ${\rm \Delta AICC}$ & $m, 10^{-2}$ & $n, 10^{-2}$ \\
            \hline
            A     &  31 & 3096.0 & 12.9 &  $2.4\pm0.1$ & 0 \\
            B     &  32 & 3083.1 & 0 &  $2.0\pm0.1$ & $4.0\pm0.3$ \\
            C     &  32 & 3331.1 & 248.0 &  $2.4\pm0.1$ & $7.6\pm0.3$ \\
            \hline
    \end{tabular}
\end{table}

\section{The exposure shift analysis}
\label{instr}

\begin{figure}
            \begin{center}
            \includegraphics[width=0.47\textwidth]{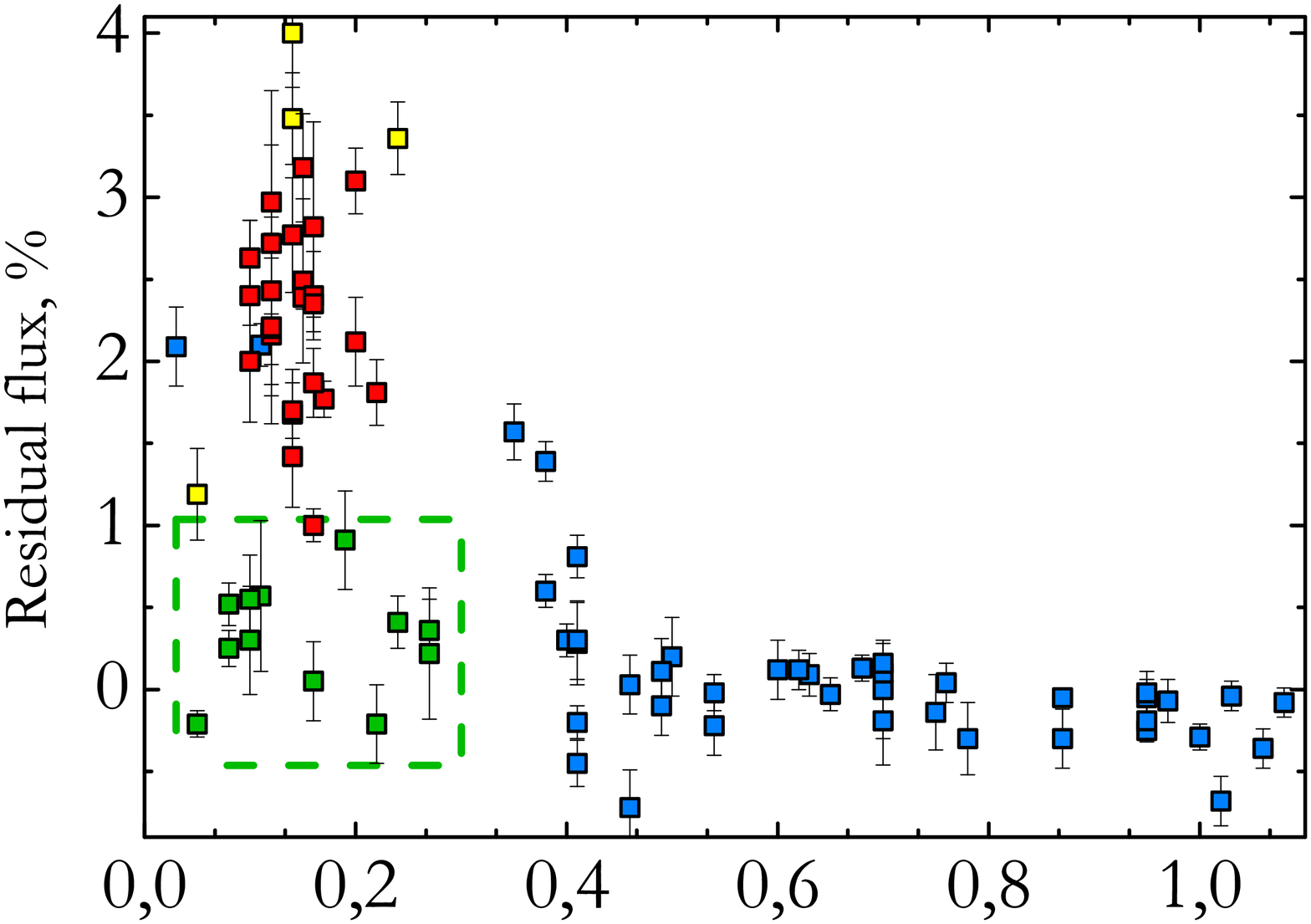}
            \caption{\rm  The dependence of  residual fluxes in the bottom of lines on the widths of the bottom of lines. Blue points correspond to all saturated Ly$\alpha$ forest lines  with $\lambda>3500$\,\AA\, and reached the zero-flux level within 5\%. The yellow points mark Ly$\alpha$ absorbers which have also Ly$\beta$ absorption lines in the spectrum. The red points represent the residual fluxes in H$_2$ lines. Among the Ly$\alpha$ forest lines there are lines with LFR $< 1$\,\% and the width of the bottom $< 0.3$\,\AA. We marked them by green colour. Existence of such lines indicate that the residual fluxes in H$_2$ lines are not the instrumental effect (see text for detail). }
            \label{AppA1}
            \end{center}
        \end{figure}

\begin{figure*}
    \begin{center}
           \includegraphics[width=0.95\textwidth]{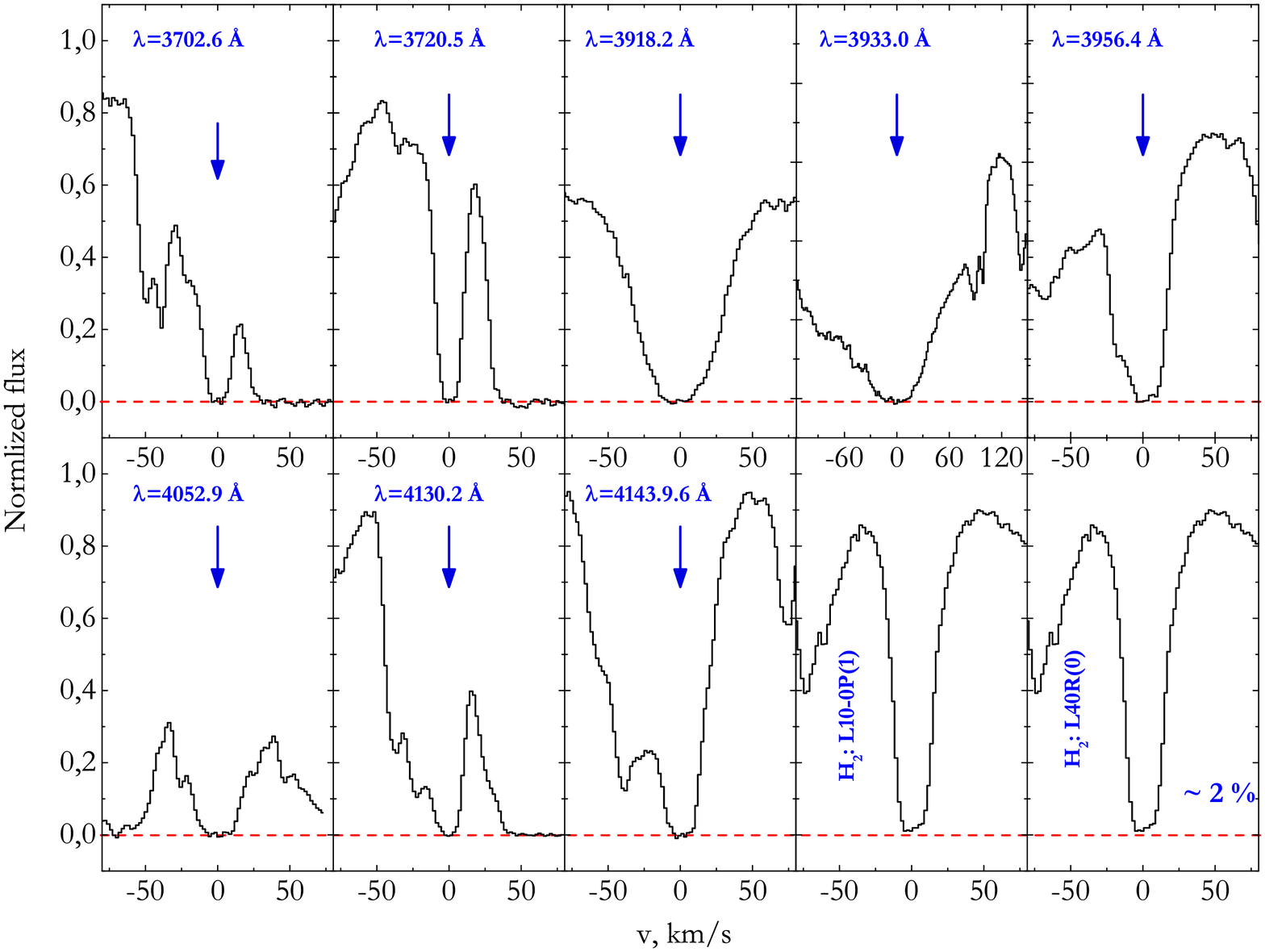}
            \caption{\rm The profiles of the Ly$\alpha$ forest lines corresponded to green filled squares in Fig.\,\ref{AppA1}.
                         For comparison we show also two H$_2$ lines, L10-0P(1) and L4-0R(0), in the two bottom right panels.
                         The red horizontal line represents the zero flux level. }
            \label{AppA2}
     \end{center}
 \end{figure*}

\begin{figure*}
    \begin{center}
           \includegraphics[width=0.95\textwidth]{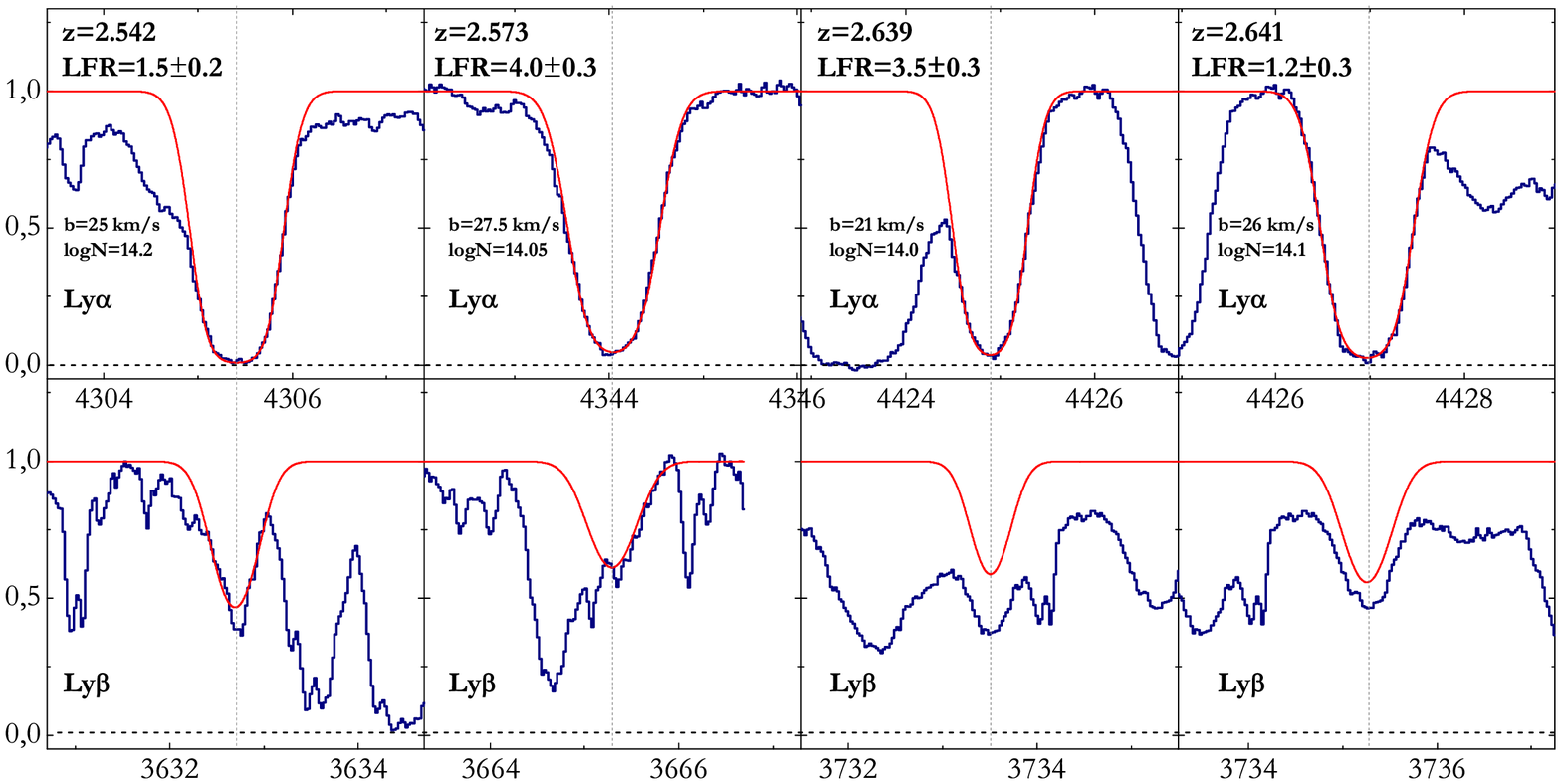}
            \caption{\rm The profiles of the Ly$\alpha$ forest lines corresponded to yellow filled squares in Fig.\,\ref{AppA1}. The top panels show Ly$\alpha$ absorptions. The bottom panels show corresponded Ly$\beta$ absorptions. The best Voigt profile fit is shown by red lines. }
            \label{AppA3}
     \end{center}
\end{figure*}

Because the final spectrum of Q\,0528$-$250 is the co-addition of 27 exposures the non-zero residual flux
at the bottom of saturated absorption lines (such as H$_2$ lines) could arise due to the average velocity
shifts between exposures, up to 500\,m\,s$^{-1}$, and/or intra-order velocity distortions, up to
1500\,m\,s$^{-1}$ (e.g. \citealt{Whitmore2010, Rahmani2013}). Also, such a small LFR could be the result
of the dramatic shift of even one exposure (e.g. \citealt{Rahmani2013}). In that case, all narrow
saturated absorption lines would have the same non-zero residual fluxes, whereas the wide saturated
absorption lines (like most of the Ly$\alpha$ forest lines) will go to zero. Fig.\,\ref{AppA1} shows the
residual fluxes obtained in absorption lines that have the residual flux within 5 per cent of the
continuum and located in the region 3500--4700\,\AA\ in comparison to their widths at the bottom. The
measurement procedure was described in Section\,\ref{Zeroflux}. The red filled squares correspond to H$_2$
lines, and the blue, green and yellow filled squares correspond to the Ly$\alpha$ forest lines. The points
marked in green represent the sample of the narrow saturated Ly$\alpha$ forest lines which look like the
H$_2$ lines, but the residual flux at the bottom of the lines goes to zero. The profiles of these lines
are shown in Fig.\,\ref{AppA2}. The existence of such lines in the spectrum of Q\,0528$-250$ implies that
the dramatic shift between exposures does not exist, otherwise the residual flux would be present in these
lines too. There are also several Ly$\alpha$ forest lines that have non-zero residual fluxes comparable or
even larger than the average H$_2$ residual flux. By performing visual inspections of these lines,
Ly$\beta$ absorption lines for four of them (yellow squares in Fig.\,\ref{AppA1})were found, due to the
fact that corresponding Ly$\alpha$ lines have redshifts large than 2.43. For systems with z lower than
2.43 the Ly$\beta$ absorption lines are located in the Lyman-break region in the spectrum. The
Fig.\ref{AppA3} shows the result of Voigt fitting analysis of these lines. All of them have small H\,{\sc
I} column densities $(\log N<14.3)$ and large Doppler parameters $({\rm b>25\,km\,s^{-1}})$.

\section{The best fit of H$_2$ lines}

Here we present the figures of the best fit of the H$_2$ lines.

\begin{figure*}
\begin{center}
       \includegraphics[width=0.95\textwidth]{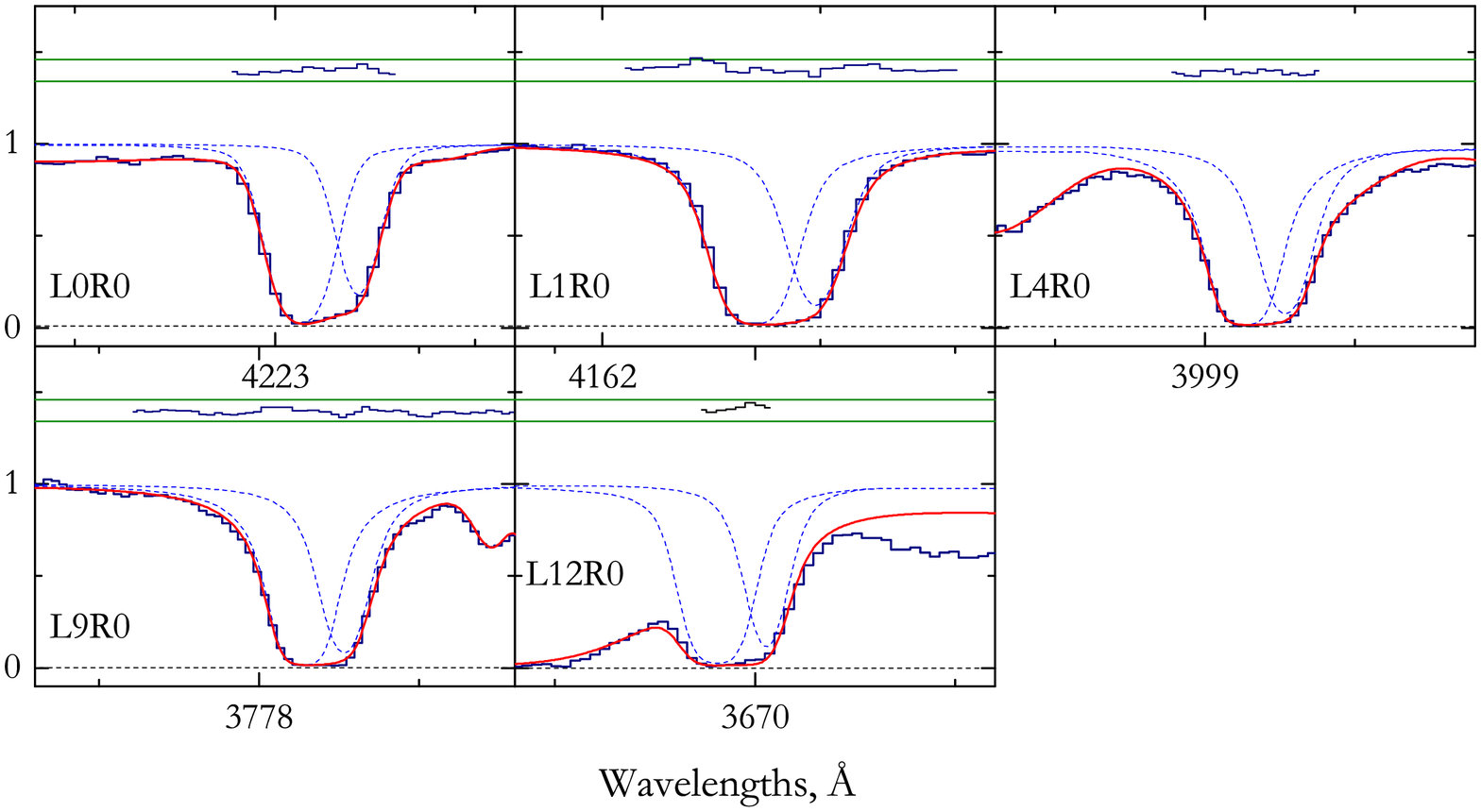}
        \caption{\rm The best fit of the H$_2$ absorption lines from ${\rm J=0}$ level at ${\rm z_{abs}=2.811}$ toward Q\,0528$-$250 (after incomporating the correction for partial coverage). Two components of absorption system are shown by blue dashed lines. The residuals between fit and data are shown in the top of each panel by dark blue line. Horizontal green lines represent $\pm3\,\sigma$ levels.}
        \label{J0lines}
\end{center}
\end{figure*}

\begin{figure*}
\begin{center}
        \includegraphics[width=0.95\textwidth]{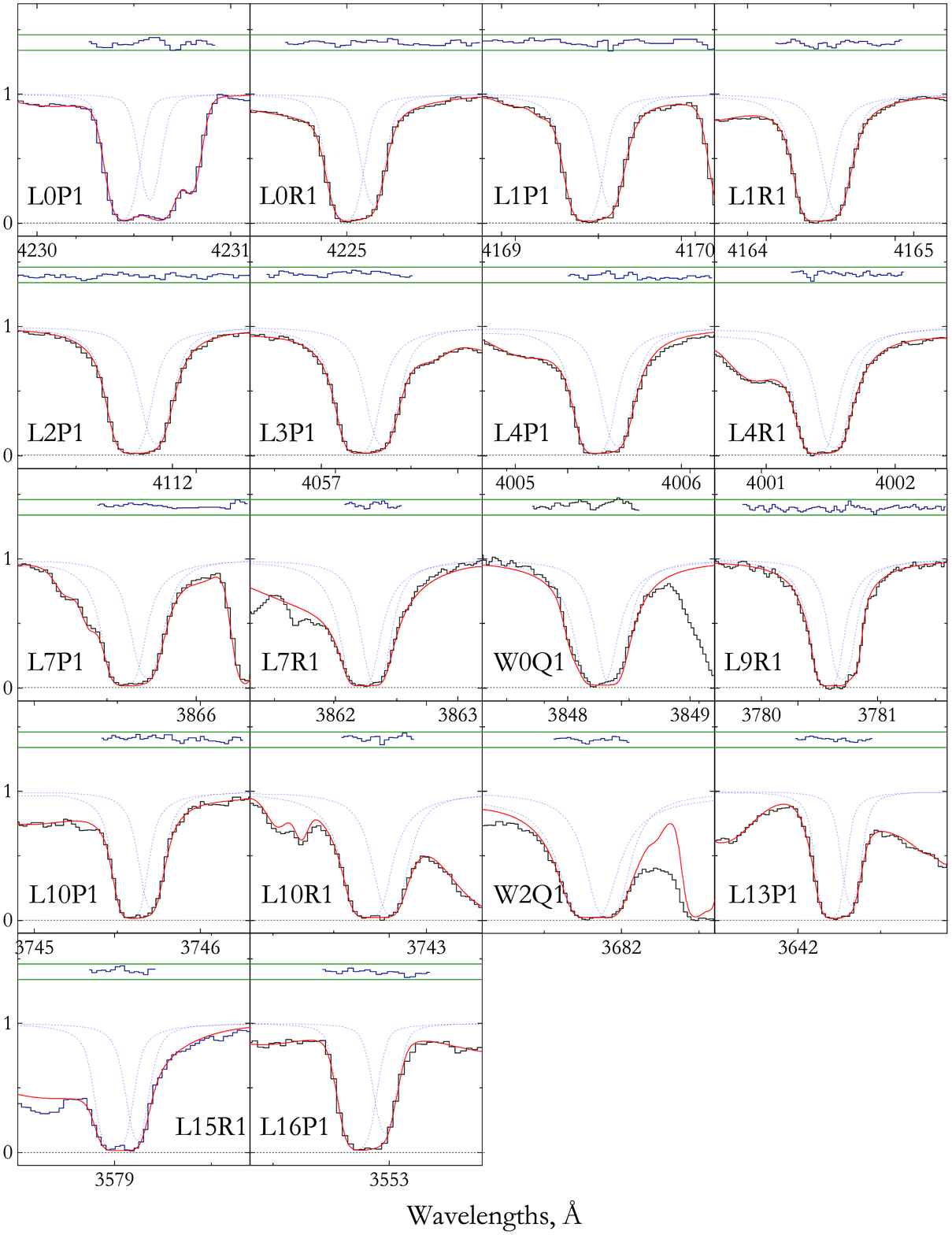}
        \caption{\rm The best fit of the H$_2$ absorption lines from ${\rm J=1}$ level at ${\rm z_{abs}=2.811}$ toward Q\,0528$-$250. Colors and lines are the same as in Figure\,\ref{J0lines}.}
        \label{J1lines}
\end{center}
\end{figure*}

\begin{figure*}
\begin{center}
        \includegraphics[width=0.95\textwidth]{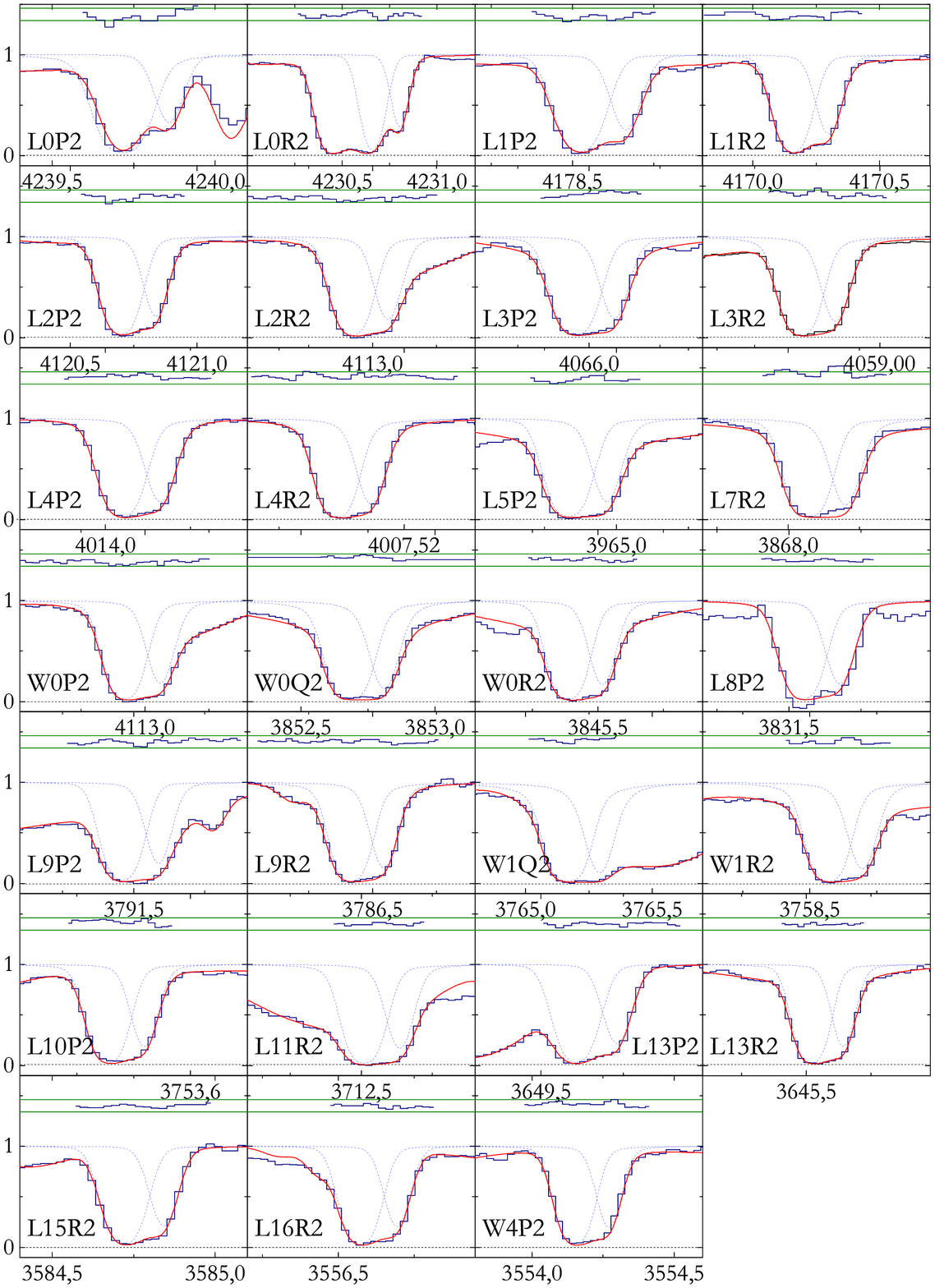}
        \caption{\rm The best fit of the H$_2$ absorption lines from ${\rm J=2}$ level at ${\rm z_{abs}=2.811}$ toward Q\,0528$-$250. Colors and lines are the same as in Figure\,\ref{J0lines}.}
        \label{J2lines}
\end{center}
\end{figure*}

\begin{figure*}
\begin{center}
        \includegraphics[width=0.9\textwidth]{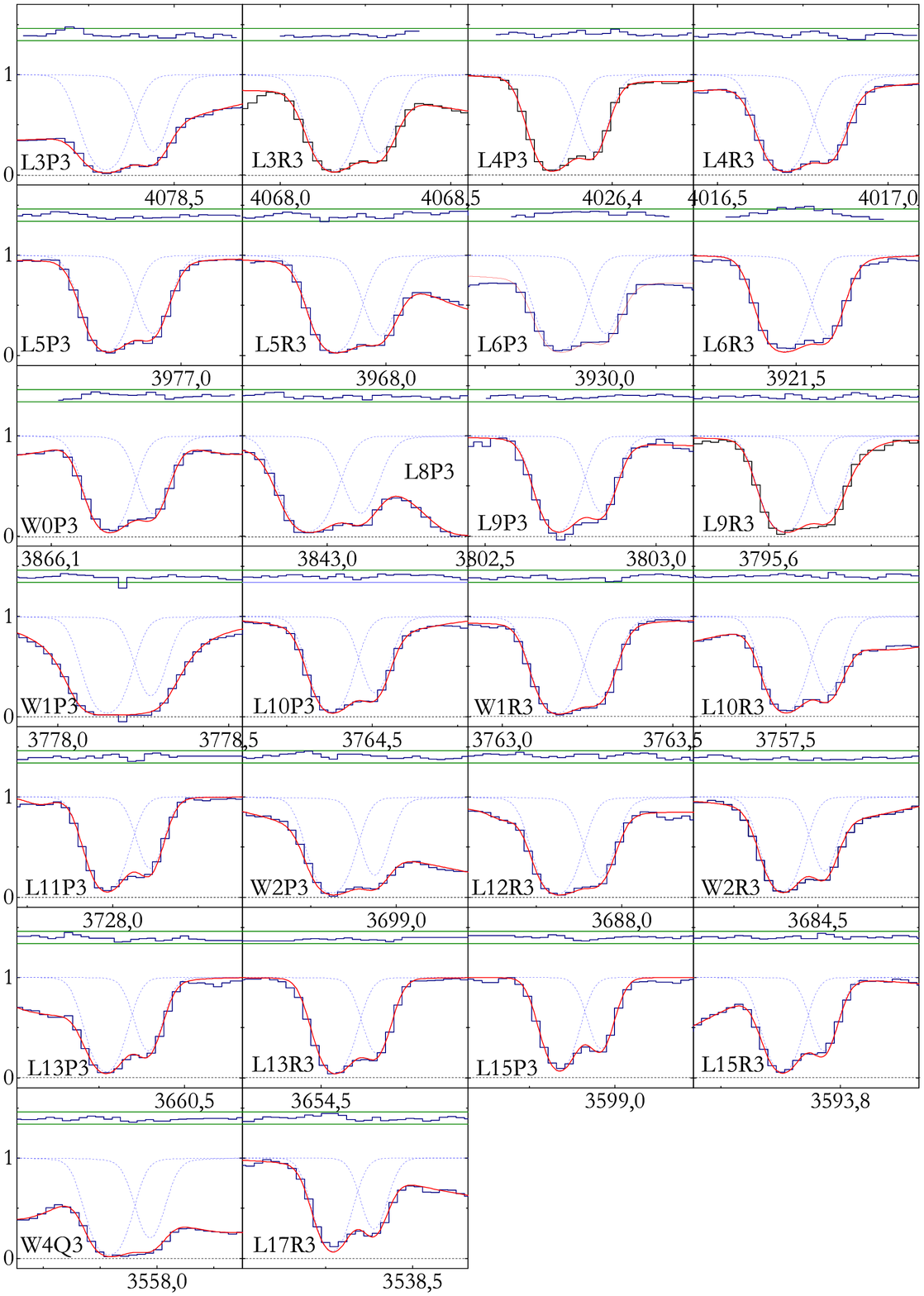}
        \caption{\rm The best fit of the H$_2$ absorption lines from ${\rm J=3}$ level at ${\rm z_{abs}=2.811}$ toward Q\,0528$-$250. Colors and lines are the same as in Figure\,\ref{J0lines}.}
        \label{J3lines}
\end{center}

\end{figure*}
\begin{figure*}
\begin{center}
       \includegraphics[width=0.95\textwidth]{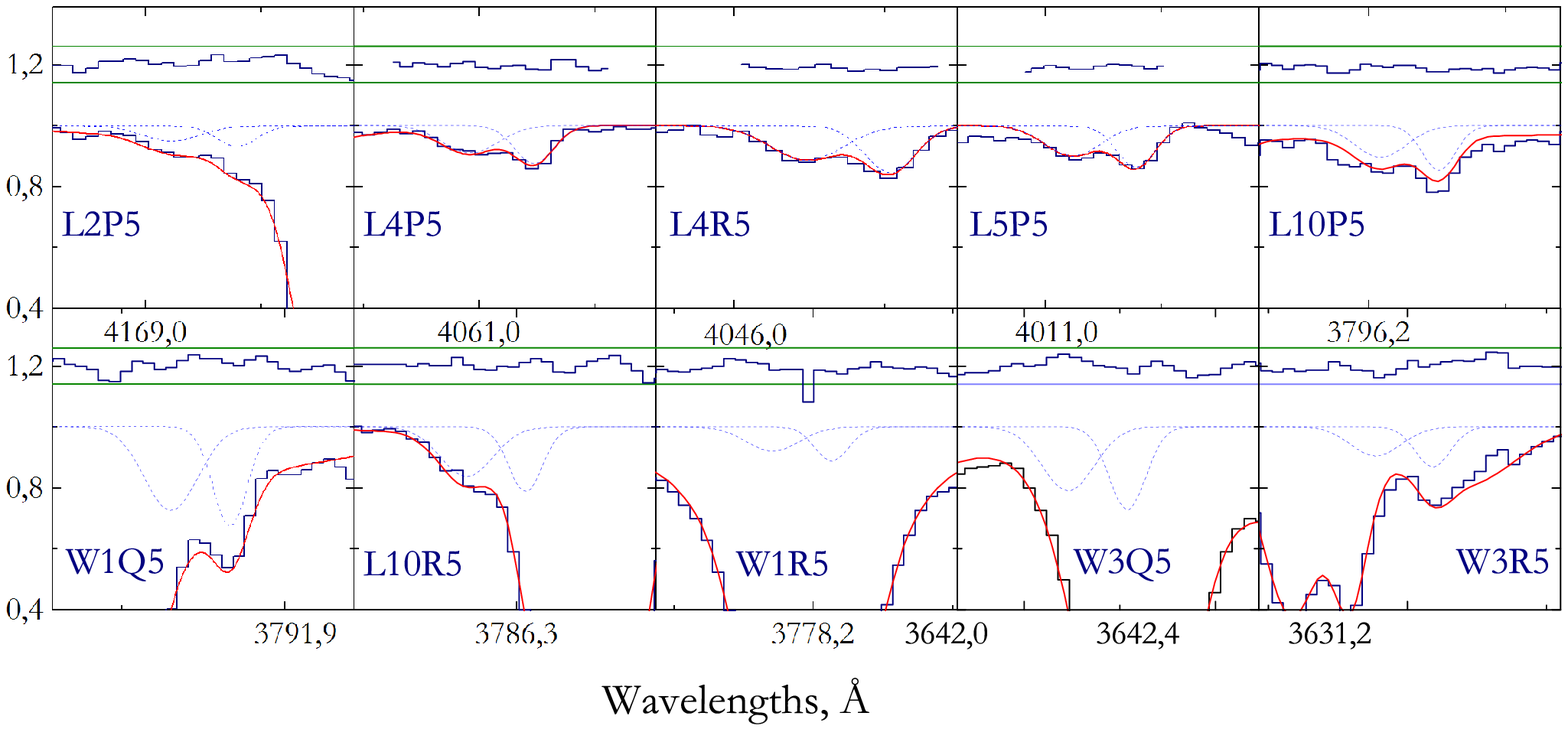}
        \caption{\rm The best fit of the H$_2$ absorption lines from ${\rm J=5}$ level at ${\rm z_{abs}=2.811}$ toward Q\,0528$-$250. Colors and lines are the same as in Figure\,\ref{J0lines}.}
        \label{J5lines}
\end{center}
\end{figure*}

\begin{figure*}
\begin{center}
        \includegraphics[width=0.95\textwidth]{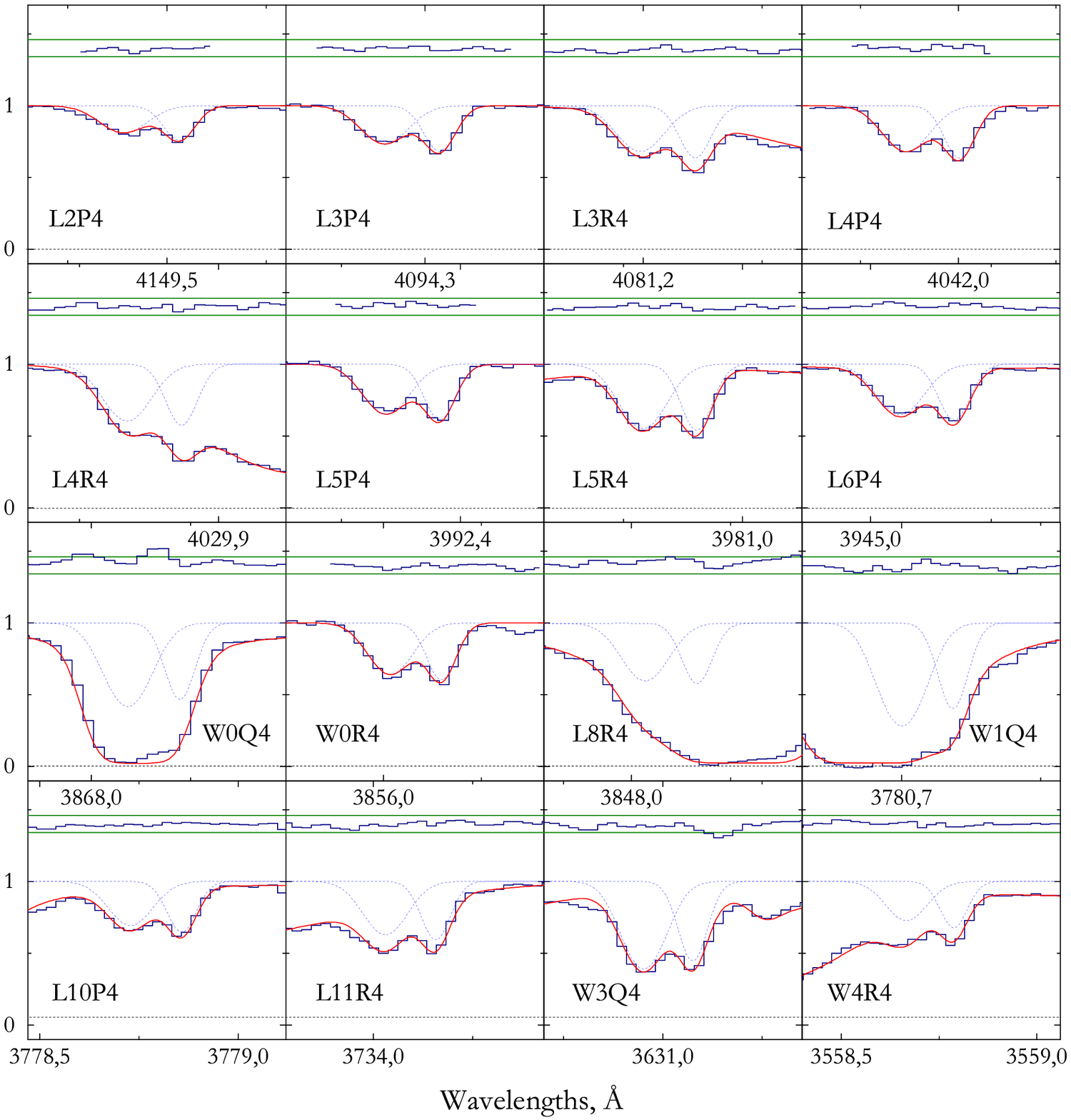}
        \caption{\rm The best fit of the H$_2$ absorption lines from ${\rm J=4}$ level at ${\rm z_{abs}=2.811}$ toward Q\,0528$-$250. Colors and lines are the same as in Figure\,\ref{J0lines}.}
        \label{J4lines}
\end{center}
\end{figure*}

\label{lastpage}

\bsp

\end{document}